\documentclass[12pt,a4paper,epsfig]{article}
\usepackage{amsfonts}
\usepackage{epsfig}

\title{{\bf On the Supersymmetric Spectra \\ of two \\ Planar Integrable Quantum Systems}}

\author{M. A. Gonzalez Leon$^{(a)}$, M. de la Torre Mayado$^{(b)}$\\ J.
Mateos Guilarte$^{(b)}$, M.J. Senosiain$^{(c)}$
\\ {\normalsize {\it $^{(a)}$ Departamento de Matematica
Aplicada and IUFFyM}, {\it Universidad de Salamanca, SPAIN}}\\ {\normalsize {\it $^{(b)}$ Departamento de Fisica and IUFFyM} ,{\it
Universidad de Salamanca, SPAIN}}\\{\normalsize{\it $^{(c)}$
Departamento de Matematicas}, {\it Universidad de Salamanca,
SPAIN}}}
\date{}

\begin{document}
\maketitle
{ \sl  \lq\lq Nil actum credens cum quid
superesset agendum" \\  \lq\lq Nothing has been done if
something remains to be done"}
\begin{abstract}
Two planar supersymmetric quantum mechanical systems built around the quantum integrable Kepler/Coulomb
and Euler/Coulomb problems are analyzed in depth. The supersymmetric spectra of both systems are unveiled,
profiting from symmetry operators not related to invariance with respect to rotations. It is shown analytically
how the first problem arises at the limit of zero distance between the centers of the second problem. It appears
that the supersymmetric modified Euler/Coulomb problem is a quasi-isospectral deformation
of the supersymmetric Kepler/Coulomb problem.
\end{abstract}
\tableofcontents
\section{Introduction}

During the last forty years a very interesting jump from symmetry to supersymmetry has taken place,
determining theoretic particle spectra in quantum field theories with
extremely appealing characteristics, see e.g. \cite{Weinberg}. Unlike many quantum
field theoretical models, the supersymmetric systems are frequently amenable
to non-perturbative treatments, see e.g. \cite{Wipf}, but the main feature is
that fermions and bosons are jointly assembled in multiplets, a fact, although
suggestive, that has not yet experimentally confirmed. Thus, mechanisms of
spontaneous supersymmetry breaking must be investigated in the
search for explanations of the apparent lack of supersymmetry in
nature. In a series of papers, Witten, \cite{Wit1}, \cite{Wit2}, and
\cite{Wit3}, proposed the analysis of this phenomenon in the simplest
possible setting: supersymmetric quantum mechanics. A new area of
research in quantum mechanics was born, with far-reaching
consequences both in mathematics and physics. The relation between
the Dirac operator in electromagnetic and/or gravitational fields -
the supercharge - with the Klein-Gordon operator - the
supersymmetric Hamiltonian - provided a guide for the building of
supersymmetric quantum mechanical systems. The factorization method
of identifying the spectra of Schrodinger operators by means of
first-order differential operators, see \cite{Inf} for a review, is
another antecedent of supersymmetric quantum mechanics that can also
be traced back to the 19th century through the Darboux transform
\cite{Darboux}. In its modern version, supersymmetric quantum
mechanics prompted the study of many one-dimensional systems from a
physical point of view. A good deal of this work can be found in
References \cite{Casal}, \cite{Ber}, \cite{Khare}, \cite{Junker}.
Several examples of this structure with emphasis in the
semi-classical behavior of non-harmonic oscillators have been worked
out in \cite{Car}.

The formalism of physical supersymmetric systems with more than one
bosonic/fermionic pairs of degrees of freedom was first developed by Andrianov, Ioffe and
coworkers in a series of papers, \cite{Ioffe}, \cite{Ioffe2},
published in the eighties.  Factorability, even though
essential in N-dimensional SUSY quantum mechanics, is not so
effective as compared with the one-dimensional situation. Some
degree of separability is also necessary to achieve analytical
results. For this reason we started a research program in the
two-dimensional supersymmetric classical mechanics of Liouville
systems, \cite{Perelomov}; i.e., those systems separable in elliptic, polar,
parabolic, or Cartesian coordinates, see papers \cite{AM} and
\cite{AoP}. We therefore follow this path in the quantum domain for Type I
Liouville models in \cite{JPA}.

Nevertheless, the authors from Saint Petersburg University mentioned above
considered from the earlier eighties higher-than-one-dimensional SUSY quantum mechanics
from the point of view of the factorization of N-dimensional quantum systems, \cite{Andr},
\cite{Andr1}. Ioffe et al. also studied the interplay between
supersymmetry and integrability in quantum and classical settings in
other types of model in References \cite{Andr2}, \cite{Can}, \cite{Can1}. In
these papers, a new structure was introduced \cite{Spiridonov}:
second-order (and higher-order) supercharges provided intertwined
scalar Hamiltonians even in the two-dimensional (and
higher-dimensional) case, see the review papers \cite{Ioffe3} and
\cite{Ioffe9}. This higher-order SUSY algebra allows for new forms
of non-conventional separability in two dimensions. There are two
possibilities: (1) a similarity transformation performs the separation
of variables in the supercharges and some eigen-functions (partial
solvability) can be found, see \cite{Ioffe4}, \cite{Ioffe5}. (2) One
of the two intertwined Hamiltonian allows for exact separability:
the spectrum of the other is known, \cite{Ioffe6}, \cite{Ioffe7}.

Our purpose in this paper is to describe planar supersymmetric
systems - two bosonic/fer\-mio\-nic pairs of degrees of freedom -
such that the Bose-Bose and Fermi-Fermi (scalar) Hamiltonians will
be separable. In Reference \cite{Eisenhart} Eisenhart classified all
the quantum systems with se\-pa\-rable Sch$\ddot{\rm o}$dinger
equations in Cartesian, polar, parabolic, and elliptic coordinates.
We shall address two planar supersymmetric separable systems, one in
polar, the other in elliptic coordinates.

We shall first consider the planar supersymmetric Kepler/Coulomb
problem showing the separability in polar coordinates. We strongly
rely on the work by Wipf et al in papers \cite{KLPW},
\cite{KLPW1} where they solve this problem in any dimension D by
finding a supersymmetric matrix Runge-Lenz vector and describing algebraically
the spectral problem in terms of the irreducible
representations of ${\mathbb S}{\mathbb O}(D+1)$. Instead we shall
attack the spectral problem in the Bose-Bose sector, finding the
bound state energies in terms of the Casimir eigenvalues of the
irreducible representations of ${\mathbb S}{\mathbb O}(3)$, whereas
the eigenfunctions are generalized Laguerre polynomials. The
scattering eigenfunctions in this sector, as well as in the
Fermi-Fermi sector, are generic confluent hypergeometric functions
given in terms of infinite series. By acting with the supercharges,
we provide in turn both the bound state and scattering
eigenfunctions in the Fermi-Bose Sector. We remark that, following a
previous work on the supersymmetric classical Kepler/Coulomb
problem, \cite{Manton}, Heumann chose another superpotential
\cite{Heumann} leading to a supersymmetric quantum mechanical system
where the Runge-Lenz vector  is no longer an invariant even in the
Bose-Bose and Fermi-Fermi sectors.

In the second half of the paper we study a supersymmetric quantum
mechanical system built from the classical Euler problem: a light
particle moving in the gravitational field created by two fixed
Newtonian centers of force restricted to the plane of the centers
\cite{OM}. Besides Euler, this system attracted the interest of investigators
 of stature such as Lagrange, Jacobi, Liouville, Darboux and others, see
\cite{OM} to read a brief history of the subject,  on a double
front: 1) because of the potential applications in celestial mechanics,
e.g., as an intermediate step in the three-body problem. 2) Because
the Euler problem was a playground where the ideas of integrability,
curvilinear coordinates, Hamilton-Jacobi separability, of such
importance in classical dynamics, were tested. All this was
imported by Pauli \cite{Pauli} to the quantum domain in his research
on the spectrum of the $H_2^+$ hydrogen ion molecule. A Chapter of
Pauling and Wilson's book on Quantum mechanics \cite{Pauling} is
devoted to the developments in this quantum problem up to the mid
thirties of the past century.

We shall address a supersymmetric quantum mechanical system such
that the scalar Hamiltonians in the Bose-Bose and Fermi-Fermi
sectors are related to the quantum mechanical Euler/Pauli
Hamiltonian. We are guided by the separability of the Schr$\ddot{\rm
o}$dinger equation in ellip\-tic coor\-di\-nates: half the sum and
half the difference of the distance to the centers. This is the main
property of the Euler-Pauli Hamiltonian allowing for its
integrability. We choose our scalar Hamiltonians fulfilling this
property but supersymmetry requires the energy to be non-negative.
We are forced to add a \lq\lq classical" piece to the EP potential
energy that pushes the ground state energy to zero. All this is
achieved by the choice of a superpotential inspired in the Ioffe/Wipf et
al superpotential for the supersymmetric Kepler/Coulomb problem
(our superpotential tends to the ABI/KLPW superpotential when the two
centers collapse). In Reference \cite{MAJM}, however, we explored
other possibilities in comparison with this superpotential.

A double change of variables to one-half of ${\rm arccosh}$ and
${\rm arccos}$ of the elliptic variables transforms the separated
spectral problem into systems of entangled Razavy \cite{Razavy1} and
Whittaker-Hill \cite{Razavy2} (three-term Hill, Razavy
trigonometric) equations. The two equations in each system are in principle
independent, but they are entangled because their parameters are
determined from the integration constants -the energy and the
separation constant- that allow one to formulate the Hamiltonian
spectral problem in terms of two separated ODE's. To the best of our
knowledge, this change of variables was formulated for the first time
in \cite{Bondar}. We shall profit from the fact that for certain
values of the parameters either the Razavy or the Whittaker-Hill
equation are algebraic quasi-exactly solvable systems {\footnote{In the non supersymmetric
problem Demkov in \cite{Demkov} has shown that there are exceptional values of the energy and the separation constant for which both equations in the pair are QES providing finite solutions.}}. More
specifically, the bound states of our system arise from values of
the energy and the separation constant that lead to Razavy equations
belonging to this class of algebraic QES potentials, see \cite{FGR}.
The algebraic QES potentials have corresponding quantum Hamiltonians
that are elements of the enveloping algebra of a finite dimensional
Lie algebra (${\rm Lie}\, \, {\mathbb S}{\mathbb L}(2,{\mathbb R})$
in many cases), admitting an invariant finite module of smooth
functions as irreducible representations \cite{Finkel}. These
potentials were first studied by Turbiner \cite{Turbiner} and then
completely classified by Gonzalez-Lopez, Kamran, and Olver
\cite{GKO1}-\cite{GKO2}. The associated (weakly) orthogonal
polynomials were analyzed in full generality in \cite{FGR0} and all
this machinery was applied to study the Razavy trigonometric
potential in \cite{FGR1}.

Unlike in the non supersymmetric case, see
\cite{Demkov}-\cite{Abramov}, the Whittaker-Hill equations
unfortunately are not QES for the values of the parameters for which
the Razavy equations are QES in the supersymmetric case. Thus, we
can only give the solutions in the form of infinite series following
the theory of Hill equations, see e.g. \cite{Ars}-\cite{Hoc}. The
case of two centers of the same strength is exceptional: instead of
Whittaker-Hill equations we encounter Mathieu equations. The
solutions can be given analytically in terms of Mathieu Cosine and
Sine functions {\footnote {All the conventions on special
functions throughout the paper will follow Reference
\cite{Abramowitz}.}}\cite{Abramowitz} and it can be explicitly
checked that at the limit where the two centers collapse the
one-center wave functions are recovered.

The organization  of the paper is as follows: after this long
Introduction in the second Section \S.2 we settle down to the framework
of ${\cal N}=2$ supersymmetric Quantum Mechanics for systems of two
degrees of freedom. Because each degree of freedom can be labeled
either as Bosonic or Fermionic, we have $2^D=4$ types of state. The
Clifford algebra of ${\mathbb R}^4$ describes this
situation perfectly and helps us to define the supercharges, the Hamiltonian
structure, and the Hilbert space of states. Section \S.3 is
fully devoted to discussing the planar supersymmetric Kepler/Coulomb
problem. The Bose-Bose bound state eigenfunctions arise as
irreducible re\-pre\-sen\-tations of the dynamical ${\mathbb
S}{\mathbb O}(3)$ symmetry associated with the Runge-Lenz vector
whereas Bose-Fermi strictly positive bound states are obtained via
the action of the $\hat{Q}^\dagger$ operator on the BB bound states
(except the zero mode). The scattering states are also describe to
unveil the whole spectrum of the supersymmetric Kepler/Coulomb
problem.  A direct analysis of the spectrum of the $4\times 4$
matrix Hamiltonian and symmetry operators is also included. In
Section \S.4 we address the same program in the supersymmetric modified
Euler/Pauli two-center system. The bound state wave functions come
from polynomial $\times$ exponential solutions of (infinite) Razavy
equations multiplied by power series solutions of related
Whittaker-Hill equations. In the case of two centers of the same
strength, the WH equations are replaced by Mathieu equations. In
Section \S.5 we show how the Kepler/Coulomb spectrum reappears at
the $d=0$ limit of the two centers. Finally, we offer a final Section
with further comments on interesting generalizations of these
classically{\footnote{Here we use the word classical in a non
physical sense, i.e., classical does not refer to a class (versus
quantal) of physical phenomena.}} integrable models.

\section{${\cal N}=2$ supersymmetric planar quantum systems}
\subsection{ ${\cal N}=2$ two-dimensional SUSY quantum mechanics}

In this type of quantum mechanical systems there are two pairs of
canonically conjugated Bosonic operators - giving the position and
momentum of the particle - that we choose in coordinate
representation:
\[
\hat{p}_k =-i\hbar\frac{\partial}{\partial x_k}\, \, \, \quad , \, \, \quad
\, \hat{x}_k=x_k \, \, , \qquad [\hat{x}_k,\hat{p}_j]=i\hbar\delta_{kj} \quad .
\]
There are also two pairs of Fermionic operators -of physical
dimensions: $[\hat{\psi}_k]=M^{-\frac{1}{2}}$ - taking care of the
Fermionic degrees of freedom of the system. The Fermi operators
satisfy anti-commutation relations of the form:
\begin{equation}
\{\hat{\psi}_k,\hat{\psi}_l\}=0=\{\hat{\psi}_k^\dagger,
\hat{\psi}_l^\dagger\}\, \, , \qquad
\{\hat{\psi}_k,\hat{\psi}_l^\dagger\}=\frac{1}{m}\delta_{kl} \, \, ,
\, \, k,l=1,2
 \, \, , \label{ferr}
 \end{equation}
showing that one operator is canonically conjugated to its adjoint operator.

The Fermionic Fock space is built from the vacuum state:
$\hat{\psi}_k|0\rangle =0 \, \, , \forall k=1,2$, the two degrees of
freedom being in Bosonic states because $|0\rangle$ is an eigenstate
of zero eigenvalue of the Fermi number operator
$\hat{N}=\sum_{k=1}^2 \, \hat{\psi}_k^\dagger\hat{\psi}_k$:
$\hat{N}|0\rangle =0|0\rangle$. The creation operators acting on
$|0\rangle$ bring the system into one-particle states
$\hat{\psi}^\dagger_k|0\rangle =|1_k\rangle $, where one of the two
degrees of freedom becomes Fermionic:
$\hat{N}|1_k\rangle=|1_k\rangle $. The two-particle state - the two
degrees of freedom in Fermionic states
$\hat{N}|1_11_2\rangle=2|1_11_2\rangle $- are then obtained in a
dual way related by Fermi statistics:
$\hat{\psi}^\dagger_2|1_1\rangle = |1_1 1_2\rangle
=-\hat{\psi}^\dagger_1|1_2\rangle=- |1_2 1_1\rangle$.

The ortho-normality relations
\begin{eqnarray*}
&& \langle 0 | 0 \rangle =  \langle 1_1 | 1_1 \rangle = \langle 1_2
| 1_2 \rangle = \langle 1_11_2 | 1_11_2 \rangle =1
\\ && \langle 1_1 | 0 \rangle   = \langle 1_2 | 0 \rangle = \langle 1_2 | 1_1 \rangle = \langle 1_11_2 | 0 \rangle =\langle 1_2 1_1 | 1_1 \rangle =\langle 1_11_2 | 1_2 \rangle =0
\end{eqnarray*}
allow us to write the more general state in this finite Fermionic Fock
space ${\cal F}={\cal F}_0\oplus{\cal F}_1\oplus{\cal F}_2 $ in the
form:
\[
|f\rangle= f_0|0\rangle + \sum_{k=1}^2 \, f_{1k} |1_k\rangle + f_2 |1_11_2 \rangle \qquad , \qquad f_0, f_{1k} , f_2
\in{\mathbb C} \quad .
\]
The supersymmetric space of states is the direct product of ${\cal
F}$ with  the Hilbert space $L^2({\mathbb R}^2)$: ${\cal S}{\cal
H}={\cal H}\otimes{\cal F}=L^2({\mathbb R}^2)\otimes{\mathbb C}^4=
{\cal S}{\cal H}_0\oplus {\cal S}{\cal H}_1\oplus{\cal S}{\cal H}_2
$. The supersymmetric wave functions read:

\begin{eqnarray*} |\Psi(x_1,x_2)\rangle&=&\, f_0(x_1,x_2)|0\rangle +\sum_{k=1}^2 f_{1k}(x_1,x_2)|1_k\rangle +
 f_{2}(x_1,x_2)|1_1 1_2\rangle \\ && \hspace{1cm} f_0(x_1,x_2), f_{1k}(x_1,x_2) , f_2(x_1,x_2)
\in L^2({\mathbb R}^2) \qquad .
\end{eqnarray*}

The standard procedure for introducing supersymmetric dynamics in this setup runs as follows: One first defines
the supercharges{\footnote{By $[\cal{O}]$ we shall denote the physical dimensions of the observable ${\cal O}$.}},
\begin{eqnarray*}
\hat{Q}&=&i\hat{\psi}_k\left(\hbar\frac{\partial}{\partial x_k}+\frac{\partial W}{\partial x_k}\right)=e^{-\frac{W}{\hbar}}\hat{Q}_0e^\frac{W}{\hbar} \quad , \quad \hat{Q}^\dagger =i\hat{\psi}_k^\dagger\left(\hbar\frac{\partial}{\partial x_k}-\frac{\partial W}{\partial x_k}\right)\\
\hat{Q}_0&=&i\hbar\hat{\psi}_k\frac{\partial}{\partial x_k}  \quad , \quad [\hat{Q}]=M^{\frac{1}{2}}L T^{-1}
\end{eqnarray*}
where
\[
W(x_1,x_2): \, \, \, {\mathbb R}^2 \, \, \, \rightarrow \, \, \, {\mathbb R}
\quad , \quad [W]=ML^2T^{-1}
\]
is a real function called the superpotential.  The supercharges are thus nilpotent first-order differential operators $\hat{Q}_o^2=0=\hat{Q}^2$ that move states between the different Fermionic sectors of the space of states:
\[
{\cal S}{\cal H}_{0}\begin{array}{c} \hat{Q}^\dagger \\ \rightleftharpoons \\ \hat{Q}\end{array} {\cal S}{\cal H}_{1} \begin{array}{c} \hat{Q}^\dagger \\ \rightleftharpoons \\ \hat{Q} \end{array} {\cal S}{\cal H}_{2} \qquad .
\]

The super-Hamiltonian is defined to be
\[
\hat{H}=\frac{1}{2}\{\hat{Q},\hat{Q}^\dagger\}=\frac{1}{2}\left(\hat{Q}\hat{Q}^\dagger+\hat{Q}^\dagger\hat{Q}\right) \quad ,
\]
which implies: $[\hat{H},\hat{Q}^\dagger]=[\hat{H},\hat{Q}]=0$.
Therefore, the transformations generated by $\hat{Q}$ and
$\hat{Q}^\dagger$ are symmetries - called supersymmetries -  of the
dynamical system and the supercharges are themselves constants of
motion. Because $[\hat{H},\hat{N}]=0$, there is a ${\mathbb
Z}_2$-grading of the dynamics given by the Klein operator ${\hat
F}=(-1)^{\hat{N}}$:
\[
\hat{F}|0\rangle = |0\rangle  \, \, \,  , \, \, \,  \hat{F}|1_11_2\rangle = |1_11_2\rangle \quad , \quad  \hat{F}|1_1\rangle = -|1_1\rangle  \, \, \,  , \, \, \, \hat{F}|1_2\rangle = -|1_2\rangle \quad .
\]
In the classification of Supersymmetric Quantum Mechanics given by Kibler et al. in \cite{Kibler} our formalism
ranks in the class of a complex super-charge, $\hat{Q}$, with an involution operator $\hat{F}^2={\mathbb I}$. It is also shown in Reference \cite{Kibler} that it is equivalent to another supersymmetric system with two real supercharges: this is the reason for the ${\cal N}=2$ in the title.

\subsection{Clifford algebra representation}

In order to skip abstract ket/bra Dirac algebra we represent the Fermi operators by means of the generators of
the Clifford algebra of ${\mathbb R}^4$:
\begin{eqnarray*}
\hat{\psi}_1&=&\frac{1}{2\sqrt{m}}\left(\gamma^1+i\gamma^3\right)\, \, \quad , \, \, \quad \quad \quad \hat{\psi}_2=\frac{1}{2\sqrt{m}}\left(\gamma^2+i\gamma^4\right)\\ \hat{\psi}_1
&=&\frac{1}{\sqrt{m}}\left(
\begin{array}{cccc}
 0 & 1 & 0 & 0  \\
 0 & 0 & 0 & 0  \\
 0 & 0 & 0 & 1  \\
 0 & 0 & 0 & 0
 \end{array}
\right) \, \, \quad , \,  \quad \hat{\psi}_2 =\frac{1}{\sqrt{m}}\left(
\begin{array}{cccc}
 0 & 0 & 1 & 0 \\
 0 & 0 & 0 &\hspace{-0.2cm}-1 \\
 0 & 0 & 0 & 0 \\
 0 & 0 & 0 & 0  \\
 \end{array}
\right) \quad .
\end{eqnarray*}
One can check that this is a minimal realization of the Fermionic anticommutation rules (\ref{ferr}) and the
Fermionic Fock space becomes the space of four-component Euclidean spinors with basis:
\[
|0\rangle \, \, \rightarrow \, \, \left(\begin{array}{c} 1 \\ 0 \\ 0 \\ 0\end{array}\right)\, \, \, , \, \, \,
|1_1\rangle \, \, \rightarrow \, \, \left(\begin{array}{c} 0 \\ 1 \\ 0 \\ 0\end{array}\right)\, \, \, , \, \, \, |1_2\rangle \, \, \rightarrow \, \, \left(\begin{array}{c} 0 \\ 0 \\ 1 \\ 0 \end{array}\right)\, \, \,  , \, \, \,
|1_11_2\rangle \, \, \rightarrow \, \, \left(\begin{array}{c} 0 \\ 0 \\ 0 \\ 1 \end{array}\right)\quad .
\]
The supercharges are $4\times 4$-matrices of differential operators
\[
\hat{Q}=\frac{i}{\sqrt{m}}\left(
\begin{array}{cccc}
 0 & D_1 & D_2 & 0 \\
 0 & 0 & 0 & \hspace{-0.2cm} -D_2  \\
 0 & 0 & 0 & D_1 \\
 0 & 0 & 0 & 0 \\
 \end{array}
\right) \, \, \, , \quad
\hat{Q}^\dagger=\frac{i}{\sqrt{m}}\left(
\begin{array}{cccc}
 0 & 0 & 0 & 0  \\
\bar{D}_1 & 0 & 0 & 0 \\
 \bar{D}_2 & 0 & 0 & 0  \\
 0 & \hspace{-0.2cm} -\bar{D}_2 & \bar{D}_1 & 0  \\
 \end{array}
\right)
\]
where $D_k=\hbar \partial_k+\frac{\partial W}{\partial x_k}$  and
$\bar{D}_k=\hbar\partial_k-\frac{\partial W}{\partial x_k}$. The
super-Hamiltonian is also a $4\times 4$-matrix of differential
operators
\[
\hat{H}=\hat{H}_0\otimes {\mathbb I}_4
-\hbar\sum_{k=1}^2\sum_{l=1}^2 \, \frac{\partial^2 W}{\partial
x_k\partial
x_l}\hat{\psi}_k^\dagger\hat{\psi}_l=\left(\begin{array}{cccc}
\hat{H}_0 & 0 & 0 & 0\\ 0 & \hat{H}_1^{11} & \hat{H}_1^{12} & 0 \\ 0
& \hat{H}_1^{21} & \hat{H}_1^{22} & 0 \\ 0 & 0 & 0 & \hat{H}_2
\end{array}\right)
\]
with a block-diagonal structure inherited from the eigen-spaces of the Fermi number operator:
\[
\hat{N}=\hat{\psi}^\dagger_1\hat{\psi}_1+\hat{\psi}^\dagger_2\hat{\psi}_2=\frac{1}{m}\left(
\begin{array}{cccc}
 0 & 0 & 0 & 0 \\
 0 & 1 & 0 & 0  \\
 0 & 0 & 1 & 0 \\
 0 & 0 & 0 & 2 \\
 \end{array} \right) \qquad .
\]
Thus, in the $\hat{F}=+1$ eigen-sectors of ${\cal S}{\cal H}$ the
Hamiltonian act by means of the scalar ordinary Schr$\ddot{\rm
o}$dinger operators:
\begin{eqnarray*}&&\hat{H}_0\equiv\left.\hat{H}\right|_{{\cal S}{\cal H}_0} = \frac{1}{2m} \left(
-\hbar^2\bigtriangleup+\partial_1W\partial_1W+\partial_2W\partial_2W+\hbar\bigtriangleup W\right)\\
&& \hat{H}_2\equiv\left.\hat{H}\right|_{{\cal S}{\cal H}_2}=\frac{1}{2m}\left(-\hbar^2\bigtriangleup+
\partial_1W\partial_1W+\partial_2W\partial_2W-\hbar\bigtriangleup W\right)\quad .\end{eqnarray*}
In ${\cal S}{\cal H}_1$, however, the super-Hamiltonian reduces to
the $2\times 2$-matrix Schr$\ddot{\rm o}$dinger  operator:
\[\hat{H}_1\equiv\left.\hat{H}\right|_{{\cal S}{\cal H}_1}=\left(\begin{array}{cc}\hat{H}_1^{11} & \hat{H}_1^{12} \\  \hat{H}_1^{21} & \hat{H}_1^{22} \end{array}\right)=\left(\begin{array}{cc}\hat{H}_0-\frac{\hbar}{m}\partial^2_1W & -\frac{\hbar}{m}\partial_1\partial_2W \\ -\frac{\hbar}{m}\partial_2\partial_1W & \hat{H}_0-\frac{\hbar}{m}\partial_2^2W \end{array}\right)\quad .
\]
We see that all the interactions come from the gradient and the second-order partial derivatives of the superpotential.

\section{The planar quantum Kepler/Coulomb problem and supersymmetry}
Our first goal in this survey is the development of this formalism encompassing the Hamiltonian of the Kepler-Coulomb problem.
\subsection{The quantum Kepler/Coulomb Hamiltonian}
We recall that the Kepler/Coulomb Hamiltonian describing the quantum dynamics of one-electron atoms is:
\[
\hat{K}=-\frac{\hbar^2}{2m}\bigtriangleup- \frac{\alpha}{r} \quad , \, \, \alpha>0 \qquad , \quad [\alpha]=ML^3T^{-2}\quad .
\]
We re-scale positions and momenta to new variables $x_k \, \rightarrow \, \frac{1}{m\alpha} x_k$, $\hat{p}_k \, \rightarrow \, m\alpha \hat{p}_k $ with dimensions of $[x_k]=M^2L^4T^{-2}$ and $[\hat{p}_k]=M^{-1}L^{-2}T$. By this token we see that the parameters $m$ (particle mass) and $\alpha^2$ (strength of the coupling) factor out in the
new Hamiltonian
\[
\hat{K} \, \rightarrow \, m\alpha^2 \hat{K}= m\alpha^2\left(-\frac{\hbar^2}{2}\bigtriangleup-\frac{1}{r}\right)\quad , \quad [\hat{K}]=M^{-2}L^{-4}T^2
\]
and their only physical r$\hat{\rm o}$le is to set the energy scale.

It is well known that this problem is superintegrable : The angular momentum -one scalar in the plane-
\[
 \hat{L}=-i\hbar \left( x_1\frac{\partial}{\partial x_2}-x_2\frac{\partial}{\partial x_1}\right)
\]
and the Runge-Lenz vector -two components in the plane-
\[
\hat{A}_1 = \frac{1}{2}\left(\hat{p}_2 \hat{L}+\hat{L} \hat{p}_2\right)-\frac{\hat{x}_1}{r} \, \, \, , \quad
\hat{A}_2= -\frac{1}{2}\left(\hat{p}_1 \hat{L}+\hat{L}\hat{p}_1\right)-\frac{\hat{x}_2}{r} \]
\[ [ \hat{L},\hat{A}_1 ]
= i\hbar \hat{A}_2 \, \, \, , \quad [ \hat{L},\hat{A}_2 ]
= -i\hbar \hat{A}_1
\]
both commute with $\hat{K}$:
\[
[\hat{K},\hat{L}]=[ \hat{K},\hat{A}_1 ]=[ \hat{K},\hat{A}_2 ]=0 \quad .
\]
We remark that in our variables the physical dimensions of these operators are $[\hat{L}]=ML^2T^{-1}$, $[\hat{A}_1]=[\hat{A}_2]=1$ and recall that they close the ${\mathbb S}{\mathbb O}(3)$ Lie algebra in the space of negative energy (bound states) eigen-functions of $\hat{K}\psi_E=E\psi_E, E<0$:
 \begin{eqnarray*}
&& \hspace{-1cm}[\hat{A}_1, \hat{A}_2] = -2i\hbar\left(\frac{\hat{p}_1^2+\hat{p}_2^2}{2}-\frac{1}{r}\right)\hat{L}= -2i\hbar\hat{K}\hat{L}\, \, \, , \qquad \hat{M}_1=\frac{1}{\sqrt{-2E}}\hat{A}_1 \, \, \, , \, \, \, \hat{M}_2=\frac{1}{\sqrt{-2E}}\hat{A}_2 \\ && \hat{M}_3=\hat{L}
\quad , \qquad [\hat{M}_a,\hat{M}_b]=i\hbar \varepsilon_{abc} \hat{M}_c \quad .
\end{eqnarray*}
Moreover, because the ${\mathbb S}{\mathbb O}(3)$ Casimir operator is
\[
\hat{C}^2=\hat{M}_1^2+\hat{M}_2^2+\hat{M}_3^2=-\frac{1}{2\hat{K}}\left(\hat{A}_1^2+\hat{A}_2^2\right)+\hat{L}^2
\]
and we have
\[
\hat{A}_1^2+\hat{A}_2^2=2\hat{K}\left(\hat{L}^2+\frac{\hbar^2}{4}\right)+1
\]
the Hamiltonian is given in terms of $\hat{C}^2$
\[
\hat{K}=-\frac{1}{2}\cdot \frac{1}{\hat{C}^2+\frac{\hbar^2}{4}}
\]
such that the ${\mathbb S}{\mathbb O}(3)$ symmetry is not N$\ddot{\rm o}$etherian but a dynamical symmetry. One finds immediately the bound state eigenvalues
\[
E_j=-\frac{2}{\hbar^2}\cdot \frac{1}{(2j+1)^2} \qquad , \qquad j\in \frac{1}{2}\otimes {\mathbb N} \quad ,
\]
which must be multiplied by $m\alpha^2$ to find the physical bound state energies. The bound state eigenfunctions, the ${\mathbb S}{\mathbb O}(3)$ irreducible representations, will be given in the next subsection.

\subsection{The supersymmetric Kepler/Coulomb Hamiltonian}

The superpotential proposed by Ioffe et al in \cite{Andr1} and Wipf et al (independently) in \cite{KLPW} to build the supersymmetric version of the Kepler/Coulomb problem is{\footnote{In \cite{Heumann} Heumann proposed another superpotential which spoils the Runge-Lenz vector conservation.}}:
\[
W(x_1,x_2)=-\frac{2}{\hbar}r=-\frac{2}{\hbar}\sqrt{x_1^2+x_2^2} \, \,  , \quad \frac{\partial W}{\partial x_k}=-\frac{2}{\hbar}\frac{x_k}{ r} \, \, , \quad \frac{\partial^2 W}{\partial x_k\partial x_l}=-\frac{2}{\hbar r}\left(\delta_{kl}-\frac{x_kx_l}{r^2}\right)\qquad .
\]
We also re-scale the Fermi operators $\hat{\psi}_k \, \rightarrow \, \frac{1}{\sqrt{m}}\hat{\psi}_k \, \, , \, \, [\hat{\psi}_k ]=1 $  to define the Kepler-Coulomb supercharge:
\[
\hat{Q}=i\left(\hbar\hat{\psi}_k\frac{\partial}{\partial x_k}-\frac{2}{\hbar}\hat{g}\right)\quad , \quad [\hat{Q}]=M^{-1}L^{-2}T
\]
where
\[
\hat{g}=-\frac{\hbar}{2}\frac{\partial W}{\partial x_k}\cdot\hat{\psi}_k=\frac{x_k}{r}\cdot\hat{\psi}_k \, \, \, , \qquad \hat{g}^2=0 \, \, \, , \qquad
\left\{\hat{g}^\dagger, \hat{g}\right\}=1 \, \, \, , \qquad \left(\hat{g}^\dagger\right)^2=0
\]
is a \lq\lq hedgehog" projection of the spin variables over the ${\mathbb R}^2$-plane. Explicitly,
\[
\hat{g}=\frac{1}{r}\left(
\begin{array}{cccc}
 0 & x_1 & x_2 & 0  \\
 0 & 0 & 0 & \hspace{-0.2cm}-x_2 \\
 0 & 0 & 0 & x_1 \\
 0 & 0 & 0 & 0
\end{array}
\right) \, \, \quad , \, \, \quad
\hat{g}^\dagger\hat{g}=\frac{1}{r^2}\left(
\begin{array}{cccc}
 0 & 0 & 0 & 0  \\
 0 & x_1^2 & x_1 x_2 & 0 \\
 0 & x_1 x_2 & x_2^2 & 0 \\
 0 & 0 & 0 & x_1^2+x_2^2
\end{array}
\right)\quad .
\]
The supersymmetric Kepler/Coulomb Hamiltonian reads:
\begin{equation}
\hat{H}= \left(-\frac{\hbar^2}{2}\bigtriangleup + \frac{2}{\hbar^2}\right){\mathbb I}_4-\frac{1}{r}\cdot\hat{X}
 \quad , \quad  \hat{X}=[\hat{g},\hat{g}^\dagger]={\mathbb I}_4-2\hat{N}+2\hat{g}^\dagger\hat{g}\qquad ,\label{hedg}
\end{equation}
whereas the scalar Schr$\ddot{\rm o}$dinger operators in the Bosonic sectors are:
\[
\hat{H}_0=-\frac{\hbar^2}{2}\bigtriangleup + \frac{2}{\hbar^2} -\frac{1}{r}=\hat{K}+\frac{2}{\hbar^2}
\quad , \quad
\hat{H}_2=-\frac{\hbar^2}{2}\bigtriangleup + \frac{2}{\hbar^2} +\frac{1}{r}=\hat{K}+\frac{2}{\hbar^2} +\frac{2}{r}\quad .
\]
Thus, $\hat{H}_0$ is exactly the Kepler/Coulomb Hamiltonian plus a constant needed to set to zero the energy of the
Bosonic ground state (zero mode); recall that supersymmetry forbids negative energy eigen-states. $\hat{H}_2$, however, is also (modulo a constant) the Kepler/Coulomb Hamiltonian for a particle of opposite electric charge, say a positron.
The force is repulsive and there will only be scattering states.

The matrix Schr$\ddot{\rm o}$dinger operator - already given in \cite{Andr1} circa 1984 - acting in the two-dimensional sub-space of the Fermionic Fock space such that $\hat{N}=1$ is:
\[
\hat{H}_1=\left(\begin{array}{ccc} -\frac{\hbar^2}{2}\bigtriangleup + \frac{2}{\hbar^2}-\frac{x_1^2-x_2^2}{r^3} & -\frac{2x_1x_2}{r^3}  \\ -\frac{2x_1x_2}{r^3} & -\frac{\hbar^2}{2}\bigtriangleup + \frac{2}{\hbar^2}+\frac{x_1^2-x_2^2}{r^3}\end{array}\right) \quad .
\]

\subsection{${\hat N}=0$ bound state eigenfunctions of ${\hat H}$}

The ${\hat N}=0$ bound state eigenfunctions of ${\hat H}$ are exactly the eigenfunctions of $\hat{H}_0$,
which are the same as the bound state eigenfunctions of $\hat{K}$ with displaced eigenvalues:
\begin{eqnarray*}
&&
\hspace{-1.2cm}\hat{H}_0\psi_j^{(0)}=\left(\frac{2}{\hbar^2}-\frac{1}{2}\cdot
\frac{1}{\hbar^2(j(j+1)+\frac{1}{4})}\right)\psi_j^{(0)}=\left(\frac{2}{\hbar^2}-\frac{1}{2}\cdot
\frac{1}{\hbar^2(j+\frac{1}{2})^2}\right) \psi_j^{(0)}\\ &&
j=\frac{n}{2}\in \frac{1}{2}\otimes{\mathbb N} \, \, \, , \quad
\hat{\tilde{H}}_0\psi_j^{(0)}=m\alpha^2 E^{(0)}_j\psi_j^{(0)}=\frac{2 m
\alpha^2}{\hbar^2}\left(1-\frac{1}{(2j+1)^2}\right)\psi_j^{(0)} \quad .
\end{eqnarray*}
Because of the dynamical ${\mathbb S}{\mathbb O}(3)$ symmetry of $\hat{H}_0$, the bound state eigen-functions,
which are degenerated in energy, form irreducible representations characterized by two integer or half-integer numbers, $j$ and $m$, providing the eigenvalues of the Casimir operator and $\hat{M}_3$ in common eigen-kets:
\[
 \hat{C}^2|j;m\rangle =\hbar^2 j(j+1)|j;m\rangle  \quad , \quad
\hat{M}_3|j;m\rangle =\hbar m|j;m\rangle \, \, \, , \, \, m:
-j, -j+1, \cdots , j-1, j \qquad .
\]
Using polar coordinates $r=+\sqrt{x_1^2+x_2^2}$, $\varphi={\rm arctan}\frac{x_2}{x_1}$ in coordinate representation the eigen-wave functions are of the form: $\psi_{jm}^{(0)}(r,\varphi)=\langle r;\varphi |j;m\rangle$. It is not difficult
to identify the highest weight eigen-wave functions in each irreducible representation. Let $\hat{M}_+=\hat{M}_1+i\hat{M}_2$ be the up-stairs ladder  operator
\[
\hat{M}_+=\hbar\left( j+\frac{1}{2}\right) \hat{A}_+ =\hbar^3 \left(
j+\frac{1}{2}\right) e^{i\varphi}\left\{i\frac{\partial^2}{\partial
r\partial\varphi}-\frac{1}{r}\frac{\partial^2}{\partial\varphi^2}-\frac{1}{2}\frac{\partial}{\partial
r}-\frac{i}{2
r}\frac{\partial}{\partial\varphi}-\frac{1}{\hbar^2}\right\}\quad ,
\]
which annihilates the highest weight state:
\[
\hat{M}_+\psi_{jj}^{(0)}(r,\varphi)=0 \, \, \, ,\quad \psi_{jj}^{(0)}(r,\varphi)=\langle r;\varphi |j;j\rangle =
f_j(r)e^{i j \varphi} \quad .
\]
This first-order ODE is easily integrated
\[
 f^\prime_j(r)=\left(\frac{j}{r}-\frac{2r}{\hbar^2(2j+1)}\right)f_j(r) \, \Rightarrow \, f_j(r)=r^j{\rm exp}\{-\frac{2r}{\hbar^2(2j+1)}\}
\]
and the normalized highest weight wave functions are:
\begin{equation}
\psi^{(0)}_{jj}(r,\varphi)=2\sqrt{\frac{2}{\pi}}\cdot \frac{u^j}{\hbar^2\sqrt{(2j+1)^3(2j)!}}\cdot e^{i j \varphi }
\cdot e^{-\frac{u}{2}} \quad , \quad u=\frac{4 r}{\hbar^2(2j+1)} \qquad . \label{hwwf}
\end{equation}
The down-stairs ladder operator $\hat{M}_-=\hat{M}_1-i\hat{M}_2$ in polar coordinates reads:
\[
\hat{M}_- =\hbar^3\left( j+\frac{1}{2}\right)
e^{-i\varphi}\left\{-i\frac{\partial^2}{\partial
r\partial\varphi}-\frac{1}{r}\frac{\partial^2}{\partial\varphi^2}-\frac{1}{2}\frac{\partial}{\partial
r}+\frac{i}{2 r}\frac{\partial}{\partial\varphi}-\frac{1}{\hbar^2}
\right\}\quad .
\]
From the Lie algebra we see that
\[
\psi_{jm}^{(0)} \propto \hat{M}_-^{j-m}\psi_{jj}^{(0)} \, \, \, ,\, \, m=j, m=j-1, \cdots , m=-j+1, m=-j \qquad , \qquad \hat{M}_-^{2j+2}\psi_{jj}^{(0)} =0
\]
and from the ansatz $\psi^{(0)}_{jm}(r,\phi)=N^{j-|m|}r^{|m|} P_{j-|m|}(r)e^{-\frac{2 r}{\hbar^2(2j+1)}}e^{im \varphi}$,  where the $N^{j-|m|}$ are normalization constants, the recurrence relations
\[
P_{j-|m|}(r)=\left[(2|m|+1)(|m|+1)-\frac{2(j+|m|+1)r}{(2j+1)\hbar^2}\right]P_{j-|m|-1}(r)+\frac{(2|m|+1)r}{2}
P^\prime_{j-|m|-1}(r)
\]
follow. Therefore, the $\hat{N}=0$ bound state eigen-functions are:
\begin{equation}
\psi_{jm}^{(0)}(r,\varphi)= \frac{2}{\hbar^2}\sqrt{\frac{2(j+|m|)!}{(2j+1)^3(j-|m|)!\pi}}\cdot\frac{u^{|m|}}{(2|m|)!}\cdot
{}_1F_1[|m|-j,2|m|+1,u]e^{i m\varphi}e^{-\frac{u}{2}} \label{bswf} \quad .
\end{equation}
In (\ref{bswf}) we have the Kummer confluent hypergeometric functions ${}_1F_1$ for those values of $a$ and $b$ such that the series
\[
{}_1F_1[a,b,z]=\sum_{k=0}^\infty \, \frac{(a)_k}{(b)_k}\cdot\frac{z^k}{k !}\qquad , \qquad (a)_k=\frac{\Gamma(a+1)}{\Gamma(a-k+1)} \, \, , \, \, \, \, (b)_k=\frac{\Gamma(b+1)}{\Gamma(b-k+1)}
\]
truncate to generalized Laguerre polynomials:
\[
L_{|m|-j}^{2|m|}(u)=\left(\begin{array}{c}j+|m| \\ j-|m|\end{array}\right){}_1F_1[|m|-j,2|m|+1,u] \quad .
\]
In sum, the $\hat{N}=0$ bound state eigen-functions of the planar
supersymmetric Kepler-Coulomb problem are organized as (degenerated in energy multiplets) irreducible representations of ${\mathbb S}{\mathbb O}(3)$ in $L^2({\mathbb R}^+\times{\mathbb S}^1)$ rather than in $L^2({\mathbb S}^2)$ (spherical harmonics).

\subsubsection{Ortho-normality and lower energy levels}
It is easy to check that the following ortho-normality relations hold:
\begin{eqnarray*}
\int_0^{2\pi}\, d\varphi \,\int_0^\infty\, rdr \, (\psi^{(0)}_{j_1m_1})^*(r,\varphi)\psi^{(0)}_{j_2m_2}(r,\varphi)=\delta_{j_1j_2}\delta_{m_1m_2} \, \, \, , \quad |j_1-j_2|=0,1,2,3, \cdots \\ \int_0^{4\pi}\, d\varphi \,\int_0^\infty\, rdr \, (\psi^{(0)}_{j_1m_1})^*(r,\varphi)\psi^{(0)}_{j_2m_2}(r,\varphi)=2\delta_{j_1j_2}\delta_{m_1m_2} \, \, , \quad |j_1-j_2|=\frac{1}{2}, \frac{3}{2}, \frac{5}{2}, \cdots  \, \, \quad .
\end{eqnarray*}
Note that in the case of one integer $j_1$ and one half-integer $j_2$ pairing it is necessary to integrate the $\varphi$ variable over $4\pi$ because of the double-valued representation.

We now offer two Tables with the lower energy multiplets, their 2D (cross-sections) and 3D plots:
\begin{table}[htdp]
\begin{center}
\begin{tabular}{|c|c|} \hline  Energy & Eigen-function \\ \hline & \\{\footnotesize $E_0^{(0)}=0$} & {\footnotesize $\psi_{00}^{(0)}(r,\varphi)=\frac{2}{ \hbar^2} \sqrt{\frac{2}{ \pi}}\
e^{-\frac{2 r}{ \hbar^2}}$} \\ \hline & \\ {\footnotesize $E_\frac{1}{2}^{(0)}=\frac{3}{2\hbar^2}$} & {\footnotesize $\left\{\begin{array}{c} \psi_{\frac{1}{2}\frac{1}{2}}^{(0)}(r,\varphi) = \frac{1}{ \hbar^3}\,  \sqrt{\frac{2}{
\pi}}\
r^{1/2} \ e^{-\frac{ r}{ \hbar^2}}\ e^{ i \frac{1}{2} \varphi} \\ \psi_{\frac{1}{2} \frac{-1}{2}}^{(0)}(r,\varphi) = - \frac{1}{ \hbar^3}\,
\sqrt{\frac{2}{ \pi}}\ r^{1/2} \ e^{-\frac{ r}{ \hbar^2}}\ e^{- i
\frac{1}{2} \varphi }\end{array}\right. $}
\\ \hline & \\{\footnotesize $E_1^{(0)}=\frac{16}{9\hbar^2}$} & {\footnotesize $\left\{\begin{array}{c} \psi_{1 1}^{(0)}(r,\varphi) = \frac{8}{9 \hbar^4 \sqrt{3 \pi}}\ r \
e^{-\frac{2 r}{3 \hbar^2}}\ e^{ i \varphi} \\ \psi_{10}^{(0)}(r,\varphi) = \frac{2}{ 9 \hbar^4} \sqrt{\frac{2}{3
\pi}}\ \ \left( 3 \hbar^2- 4 r \right)\
e^{-\frac{2 r}{3 \hbar^2}} \\ \psi_{1 -1}^{(0)}(r,\varphi) = \frac{8}{9 \hbar^4 \sqrt{3 \pi}}\ r \
e^{-\frac{2 r}{3 \hbar^2}}\ e^{ -i \varphi}\end{array}\right.$} \\ \hline & \\ {\footnotesize $E_\frac{3}{2}^{(0)}=\frac{15}{8\hbar^2}$} & {\footnotesize $\left\{\begin{array}{c} \psi_{\frac{3}{2} \frac{3}{2}}^{(0)}(r,\varphi) = \frac{1}{4 \hbar^5 \sqrt{3 \pi}}\
r^{3/2} \ e^{-\frac{ r}{ 2 \hbar^2}}\ e^{ i \frac{3}{2} \varphi}  \\ \psi_{\frac{3}{2} \frac{1}{2}}^{(0)}= - \frac{1}{4 \hbar^5 \sqrt{ \pi}}\
r^{1/2} \  (r- 2\hbar^2) \ e^{-\frac{ r}{2 \hbar^2}}\ e^{ i \frac{1}{2} \varphi } \\ \psi_{\frac{3}{2} \frac{-1}{2}}^{(0)}=\frac{1}{4 \hbar^5 \sqrt{ \pi}}\
r^{1/2} \  (r- 2\hbar^2) \ e^{-\frac{ r}{2 \hbar^2}}\ e^{- i \frac{1}{2} \varphi } \\ \psi_{\frac{3}{2} \frac{-3}{2}}^{(0)}=- \frac{1}{4 \hbar^5 \sqrt{3 \pi}}\
r^{3/2} \ e^{-\frac{ r}{ 2 \hbar^2}}\ e^{ - i \frac{3}{2} \varphi}\end{array}\right. $}\\ \hline & \\{\footnotesize $E_2^{(0)}=\frac{48}{25\hbar^2}$} & {\footnotesize $\left\{\begin{array}{c} \psi_{2 2}^{(0)}(r,\varphi) = \frac{16}{125\, \hbar^6\, \sqrt{15 \pi}}\
r^2\
e^{-\frac{2 r}{5 \hbar^2}}\ e^{2 i \varphi}\\ \psi_{21}^{(0)}(r,\varphi) = \frac{8}{ 125\, \hbar^6 \, \sqrt{15 \pi}}\
r \ \left( 15\, \hbar^2 - 4\, r \right)\
e^{-\frac{2 r}{5 \hbar^2}}\ e^{ i \varphi} \\ \psi_{20}^{(0)}(r,\varphi)= \frac{2}{125\, \hbar^6}\, \sqrt{\frac{2}{5
\pi}}\ \left( 25\, \hbar^4 - 40\, \hbar^2\, r  + 8\, r^2 \right)
\ e^{-\frac{2 r}{5 \hbar^2}}\\ \psi_{2 -1}^{(0)}(r,\varphi) = \frac{8}{125 \hbar^6 \sqrt{15 \pi}}\
 r\left( 15\, \hbar^2 - 4\, r \right)
\
e^{-\frac{2 r}{5 \hbar^2}}\ e^{- i \varphi}\\ \psi_{2 -2}^{(0)}(r,\varphi)=\frac{16}{125 \hbar^6 \sqrt{15 \pi}}\
r^2\ e^{-\frac{2 r}{5 \hbar^2}}\ e^{- 2 i \varphi}\end{array}\right.$}  \\
\hline
\end{tabular}
\end{center}
\end{table}
\begin{table}[htdp]
\begin{center}
\caption{3D Plots and cross-sections of the probability densities: $\hat{N}=0$.}
\begin{tabular}{|c|cccc|}  \hline  & & & & \\
{\footnotesize{Probability}} & $|\psi_{00}^{(0)}(x_1,x_2)|^2$  &
$|\psi_{\frac{1}{2}\pm \frac{1}{2}}^{(0)}(x_1,x_2)|^2$ & $|\psi_{1\pm
1}^{(0)}(x_1,x_2)|^2$ & $|\psi_{1 0}^{(0)}(x_1,x_2)|^2$ \\
{\footnotesize{density}} & & & &
\\ \hline & & & & \\  $\hbar=1$ &
\includegraphics[height=2.2cm]{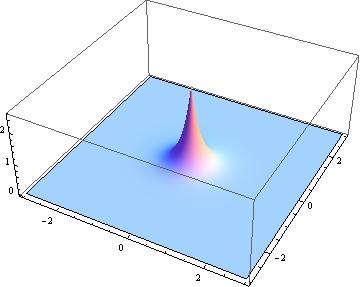} &
\includegraphics[height=2.2cm]{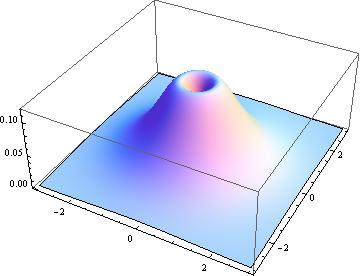} &
\includegraphics[height=2.2cm]{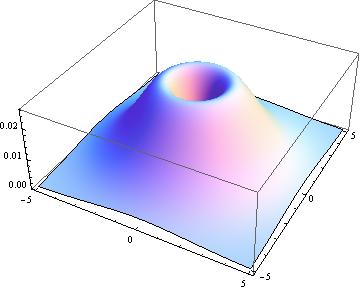} &
\includegraphics[height=2.2cm]{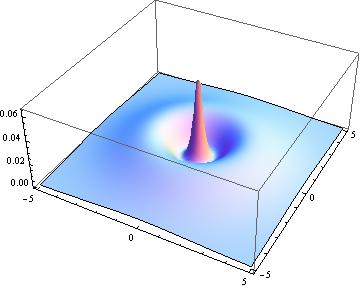}  \\ & & & & \\
 \hline & & & & \\ $\hbar=1$ &
\includegraphics[height=1.9cm]{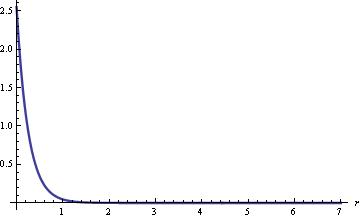}&
\includegraphics[height=1.9cm]{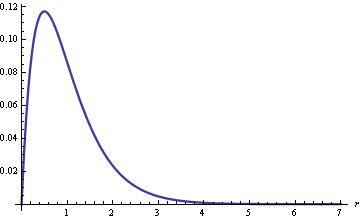} &
\includegraphics[height=1.9cm]{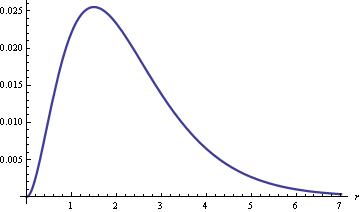} &
\includegraphics[height=1.9cm]{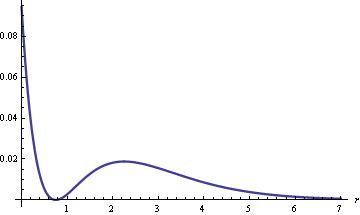}  \\ & & & & \\
 \hline
 \end{tabular}
\end{center}
\end{table}

\subsection{${\hat N}=1$ bound state eigenfunctions of ${\hat H}$}
 The supersymmetric partner eigen-states belonging to ${\cal S}{\cal H}_1$ with the same energies are of the form:
\begin{eqnarray*}
  E^{(1)}_j&=&\frac{2}{\hbar^2}\left(1-\frac{1}{(2j+1)^2}\right)\in(0,\frac{2}{\hbar^2})\, \, , \, \, j>0 \, , \\ \,
    \psi^{(1)}_{jm}(x_1,x_2)&=&\hbar\hat{Q}^\dagger \psi^{(0)}_{jm}(x_1,x_2)=\left( \begin{array}{c} 0 \\
    \psi_{jm}^{(1)(1)}(r, \varphi) \\
\psi_{jm}^{(1)(2)}(r, \varphi) \\ 0
\end{array} \right)
\end{eqnarray*}
Therefore,
\begin{eqnarray*}
(\psi_{j_1 m_1}^{(1)}(r, \varphi))^{\dag} \, \psi_{j_2m_2}^{(1)}(r, \varphi) &=&
(\psi_{j_1m_1}^{(0)}(r, \varphi))^{\dag}\, \hbar^2\, \hat{Q} \hat{Q}^{\dag}
\, \psi_{j_2m_2}^{(0)}(r, \varphi)
 \\
&=& 2\, \hbar^2\, (\psi_{j_1m_1}^{(0)}(r, \varphi))^{\dag}\, \, \hat{H} \, \,
\psi_{j_2m_2}^{(0)}(r,
\varphi)
= 2\, \hbar^2\, E_{j_2}^{(0)}\, (\psi_{j_1m_1}^{(0)}(r, \varphi))^{\dag} \,
\psi_{j_2m_2}^{(0)}(r, \varphi)
\end{eqnarray*}
because $\hat{Q}\,  \psi_{jm}^{(0)}(r, \varphi) = 0$  and $\hat{H} \,
\psi_{jm}^{(0)}(r, \varphi)  = E_j^{(0)} \,  \psi_{jm}^{(0)}(r, \varphi)$. Integration over ${\mathbb R}^2$ gives
\begin{eqnarray*}
\hspace{-0.4cm}\int_{0}^{2\pi}\, d \varphi \, \int_{0}^{\infty} r \, d r \,
(\psi_{j_1m_1}^{(1)}(r, \varphi))^{\dag} \ \psi_{j_2m_2}^{(1)}(r, \varphi) &=& 2\,
\hbar^2\, E_{j_2}^{(0)}\, \int_{0}^{2\pi}\, d \varphi \, \int_{0}^{\infty} r
\, d r \, (\psi_{j_1m_1}^{(0)}(r, \varphi))^{\dag} \, \psi_{j_2m_2}^{(0)}(r, \varphi)
\\  &=& 2\, \hbar^2\, E_{j_2}^{(1)}\, \delta_{j_1j_2}\delta_{m_1m_2} \qquad .
\end{eqnarray*}
The normalized eigenspinors $\psi_{jm}^{(1)}(r, \varphi) \rightarrow \frac{1}{\sqrt{2\, \hbar\, E_j^{(1)}}}\, \psi_{jm}^{(1)}(r,
\varphi)$
\begin{eqnarray*}
&& \psi_{j m}^{(1)(1)}(r, \varphi) = i \frac{2}{ \hbar^2} \,  \sqrt{
\frac{ 8 (j + |m|)!}{ (2j+1)^3 ((2 j +1)^2 -1) (j -|m|)! \pi}} \,
\frac{u^{|m|}}{(2 |m|)!}\, e^{i m \varphi} \, e^{-\frac{u}{2}}\\
&& \left[  {\rm cos} \varphi \, \left\{ \frac{-j+|m|}{2 |m|+1} \, \
_1F_1\left[ -j+|m|+1, 2|m|+2,u \right]  + j \, \
_1F_1\left[ -j+|m|, 2|m|+1, u \right]  \right\} \right. \\
&& \left. + \, \frac{m}{u} \, e^{-i \varphi}\, \ _1F_1\left[
-j+|m|, 2|m|+1,u \right]  \right]
\end{eqnarray*}

\begin{eqnarray*}
&& \psi_{j m}^{(1)(2)}(r, \varphi) = i \frac{2}{ \hbar^2} \,  \sqrt{
\frac{ 8 (j + |m|)!}{ (2j+1)^3 ((2 j +1)^2 -1) (j -|m|)! \pi}} \,
\frac{u^{|m|}}{(2 |m|)!}\, e^{i m \varphi} \, e^{-\frac{u}{2}}\\
&& \left[  {\rm sin} \varphi \, \left\{ \frac{-j+|m|}{2 |m|+1} \, \
_1F_1\left[ -j+|m|+1, 2|m|+2,u \right]  + j \, \
_1F_1\left[ -j+|m|, 2|m|+1, u \right]  \right\} \right. \\
&& \left. + i  \, \frac{m}{u} \, e^{-i \varphi}\, \ _1F_1\left[
-j+|m|, 2|m|+1,u \right]  \right]
\end{eqnarray*}
satisfy the spectral condition $
\hat{H} \, \psi_{jm}^{(1)}(r, \varphi) = E_j^{(1)} \, \psi_{jm}^{(1)}(r, \varphi)
\equiv \hat{H}_{1}  \ \psi_{jm}^{(1)}(r, \varphi)$
and form an orthonormal basis in ${\cal S}{\cal H}_1$.

Specifically, these two-component wave functions are linear combinations of two contiguous generalized Laguerre polynomials. The reason is that $\hat{Q}$ does not commute with the generators of the ${\mathbb S}{\mathbb O}(3)$
symmetry: $[{\hat M}_a,\hat{Q}^\dagger]\neq 0, \forall a=1,2,3 $. Therefore, the $\hat{Q}^\dagger$ action does not respect the ${\mathbb S}{\mathbb O}(3)$ irreducible representations. Nevertheless, the spinorial wave functions are characterized by the quantum numbers $j$ and $m$, although the degenerated multiplets do not form irreducible representations of ${\mathbb S}{\mathbb O}(3)$. We show next the lower spinorial probability densities:
\begin{table}[htdp]
\begin{center}
\caption{3D Plots and cross-sections of the probability densities: $\hat{N}=1$.}
\begin{tabular}{|c|cccc|}  \hline  & & & & \\
{\footnotesize{Probability}} & $|\psi_{10}^{(1)}(x_1,x_2)|^2$ &
$|\psi_{1 \pm 1}^{(1)}(x_1,x_2)|^2$ &$r|\psi_{\frac{1}{2} \pm
\frac{1}{2}}^{(1)}(x_1,x_2)|^2$ & $|\psi_{\frac{3}{2}
\pm\frac{3}{2}}^{(1)}(x_1,x_2)|^2$ \\
{\footnotesize{density}} & & & &
\\ \hline & & & & \\  $\hbar=1$ &
\includegraphics[height=2.2cm]{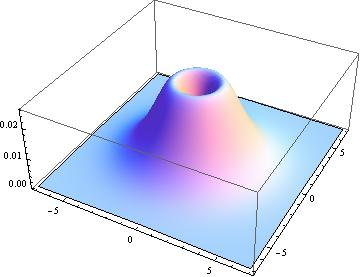} &
\includegraphics[height=2.2cm]{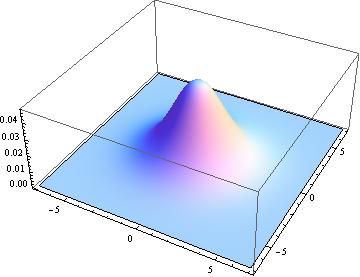} &
\includegraphics[height=2.2cm]{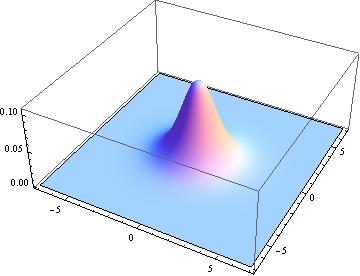} &
\includegraphics[height=2.2cm]{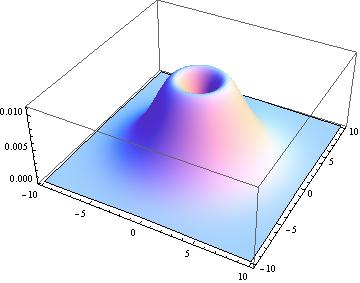}  \\ & & & & \\
 \hline & & & & \\ $\hbar=1$ &
\includegraphics[height=1.9cm]{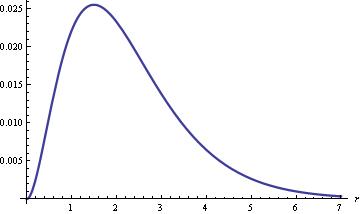}&
\includegraphics[height=1.9cm]{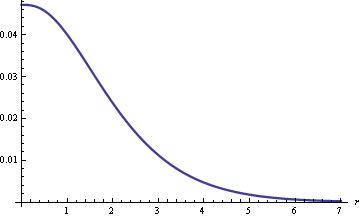} &
\includegraphics[height=1.9cm]{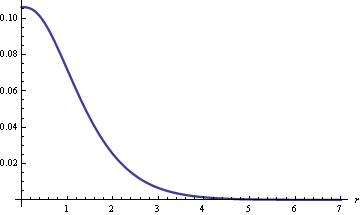} &
\includegraphics[height=1.9cm]{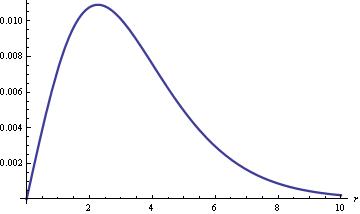}  \\ & & & & \\
 \hline
 \end{tabular}
\end{center}
\end{table}

\subsection{Scattering states and supersymmetric Hodge spectral decomposition}

On positive energy eigen-functions of the Kepler-Coulomb Hamiltonian $\hat{K}$ the normalized components of the Runge-Lenz vector and the angular momentum close the ${\mathbb S}{\mathbb O}(2,1)$ Lie algebra:
\begin{eqnarray*}
&& \hat{M}_1=-\frac{i}{\sqrt{2E}}\hat{A}_1 \, \, \, , \, \, \, \hat{M}_2=-\frac{i}{\sqrt{2E}}\hat{A}_2 \quad , \quad \hat{M}_3=\hat{L} \\ && [\hat{M}_1,\hat{M}_2]=-i \hbar \hat{M}_3 \quad , \quad [\hat{M}_3,\hat{M}_1]=i\hbar {\hat M}_2 \quad , \quad [\hat{M}_2,\hat{M}_3]=i\hbar {\hat M}_1 \quad .
\end{eqnarray*}
To search for the scattering wave functions in ${\cal S}{\cal H}_0$, the eigenfunctions of $\hat{H}_0$
\[
\left(-\frac{\hbar^2}{2}\bigtriangleup +\frac{2}{\hbar^2}-\frac{1}{r}\right)\psi_{E^{(0)}}^{(0)}(x_1,x_2)=E^{(0)}\psi_{E^{(0)}}^{(0)}(x_1,x_2) \quad , \quad E^{(0)}>\frac{2}{\hbar^2}\quad ,
\]
we profit from the fact that the spectral problem is separable into polar coordinates. The ansatz $\psi_{E^{(0)}}^{(0)}(r,\varphi)=f_{E^{(0)}}^{(0)}(r)e^{i m\varphi}$ leads to the ordinary differential equation
\[
\frac{d^2f_{E^{(0)}}^{(0)}}{dr^2}+\frac{1}{r}\frac{d f_{E^{(0)}}^{(0)}}{dr}-\frac{m^2}{r^2}f_{E^{(0)}}^{(0)}(r)+\frac{2}{\hbar^2}\left(\frac{1}{r}+E^{(0)}-\frac{2}{\hbar^2}\right)f_{E^{(0)}}^{(0)}(r)=0 \qquad .
\]
Now defining the non-dimensional variable $u=\frac{2\sqrt{2|\frac{2}{\hbar^2}- E^{(0)}|}}{\hbar}r$ we find the Bosonic
scattering solutions
\begin{equation}
f_{E^{(0)}}^{(0)}(r)=N(E^{(0)})e^{-i\frac{u}{2}}u^{|m|}{}_1F_1\left[
|m|-\frac{1}{2}-\frac{i}{\hbar\sqrt{2|\frac{2}{\hbar^2}-E^{(0)}|}},1+2|m|,i
u\right]  \label{scatsb}
\end{equation}
in terms of Kummer confluent hypergeometric functions.

Simili modo, we identify the scattering wave functions in ${\cal S}{\cal H}_2$, the eigenfunctions of ${\hat H}_2$:
\begin{eqnarray*}
&&\left(-\frac{\hbar^2}{2}\bigtriangleup +\frac{2}{\hbar^2}+\frac{1}{r}\right)\psi_{E^{(2)}}^{(2)}(x_1,x_2)=E^{(2)}\psi_{E^{(2)}}^{(2)}(x_1,x_2) \quad , \quad E^{(2)} >\frac{2}{\hbar^2} \\ &&
\frac{d^2f_{E^{(2)}}^{(2)}}{dr^2}+\frac{1}{r}\frac{d f_{E^{(2)}}^{(2)}}{dr}-\frac{m^2}{r^2}f_{E^{(2)}}^{(2)}(r)+\frac{2}{\hbar^2}\left(-\frac{1}{r}+E^{(2)}-\frac{2}{\hbar^2}\right)f_{E^{(2)}}^{(2)}(r)=0\\
&& f_{E^{(2)}}^{(2)}(r)=N(E^{(2)})e^{-i\frac{u}{2}}u^{|m|}{}_1F_1\left[
|m|-\frac{1}{2}+\frac{i}{\hbar\sqrt{2|\frac{2}{\hbar^2}-E^{(2)}|}},1+2|m|,i
u\right] \quad .
\end{eqnarray*}
The potential being repulsive, there are no bound states in ${\cal S}{\cal H}_2$.

The supersymmetry algebra now allows us to identify all the solutions of the supersymmetric spectral problem
$\hat{H}\psi_E=E\psi_E$ from the eigenfunctions of ${\hat H}_0$ and $\hat{H}_2$ with non-zero eigenvalue. The key observation is that there are two kinds of non-zero (strictly positive energy) eigenfunctions $\psi_E^\rightarrow=\hat{Q}^\dagger\psi_E\in\hat{Q}^\dagger{\cal S}{\cal H}$ and $\psi_E^\leftarrow=\hat{Q}\psi_E\in\hat{Q}{\cal S}{\cal H}$ because, if $E>0$:
\[
\hat{H}\hat{Q}^\dagger\psi_E=\hat{Q}^\dagger\hat{H}\psi_E=E\hat{Q}^\dagger\psi_E \, \, \, , \, \, \, \hat{H}\hat{Q}\psi_E=\hat{Q}\hat{H}\psi_E=E\hat{Q}\psi_E \quad .
\]
The structure of the spectrum is as follows:

\begin{itemize}

\item Ground states.

There is a unique ground state -that belongs to ${\cal S}{\cal H}_0$ and hence Bosonic-  of zero energy $E^{(0)}_0=0$:
$\psi^{(0)}_{00}(x_1,x_2)\in{\rm Ker}\hat{H}$.

\item There exist $\hat{Q}^\dagger$-exact eigenstates of three types

\begin{enumerate}
\item $\hat{Q}^\dagger$-exact - henceforth, living in ${\hat Q}^\dagger{\cal S}{\cal H}$ - bound state eigen-spinors that belong to ${\cal S}{\cal H}_1$:
 \[
 \psi^{(1)}_{E^{(1)}_j}(x_1,x_2)=\hat{Q}^\dagger \psi^{(0)}_{jm}(x_1,x_2)
 \]

\item $\hat{Q}^\dagger$-exact - henceforth, living in ${\hat Q}^\dagger{\cal S}{\cal H}$ - scattering eigen-spinors that belong to ${\cal S}{\cal H}_1$:
 \[
 E^{(1)}_-=E^{(0)} > \frac{2}{\hbar^2} \quad , \qquad \psi_{E^{(1)}_-}^{(1)}(x_1,x_2)=\hat{Q}^\dagger \psi_{E^{(0)}}^{(0)}(x_1,x_2)
 \]

\item $\hat{Q}^\dagger$-exact - henceforth, living in ${\hat Q}^\dagger{\cal S}{\cal H}$ - scattering wave-functions that belong to ${\cal S}{\cal H}_2$:
 \[
 E^{(2)}=E^{(1)}_+ > \frac{2}{\hbar^2} \quad , \qquad \psi_{E^{(2)}}^{(2)}(x_1,x_2)=\hat{Q}^\dagger \psi_{E^{(1)}_+}^{(1)}(x_1,x_2)
 \]
\end{enumerate}
\item There exist $\hat{Q}$-exact eigenstates also of three types
\begin{enumerate}
\item $\hat{Q}$-exact bound states -henceforth belonging to $\hat{Q}{\cal S}{\cal H}$-
but living in ${\cal S}{\cal H}_0$:
\[
  E^{(0)}_j=E^{(1)}_j \, \, , \, \, j>0 \, , \\ \,  \psi^{(0)}_{jm}(x_1,x_2)=\hat{Q}\psi^{(1)}_{E^{(1)}_j}(x_1,x_2)\quad .
\]
\item $\hat{Q}$-exact - henceforth, living in ${\hat Q}{\cal S}{\cal H}$ - scattering wave-functions that belong to ${\cal S}{\cal H}_0$:
 \[
 E^{(0)}=E^{(1)}_-> \frac{2}{\hbar^2} \quad , \qquad \psi_{E^{(0)}}^{(0)}(x_1,x_2)=\hat{Q} \psi_{E^{(1)}_-}^{(1)}(x_1,x_2)
 \]

\item $\hat{Q}$-exact - henceforth, living in ${\hat Q}^\dagger{\cal S}{\cal H}$ - scattering eigen-spinors that belong to ${\cal S}{\cal H}_1$:
 \[
 E^{(1)}_+=E^{(2)} > \frac{2}{\hbar^2} \quad , \qquad \psi_{E^{(1)}_+}^{(1)}(x_1,x_2)=\hat{Q} \psi_{E^{(2)}}^{(2)}(x_1,x_2)
 \]
\end{enumerate}
\end{itemize}
Because the eigenfunctions form a total set in each sub-space we have the decomposition \`{a} la
Hodge of the supersymmetric space of states:
\[
{\cal S}{\cal H}=\hat{Q}{\cal S}{\cal H}\bigoplus\hat{Q}^\dagger{\cal S}{\cal H}\bigoplus{\rm Ker}\hat{H} \quad .
\]

\begin{center}
\begin{figure}[ht]
\includegraphics[height=6cm]{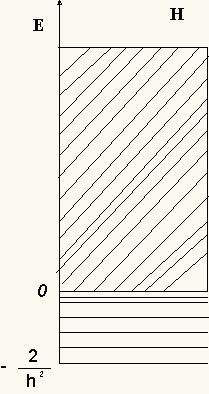}\hspace{1cm} \includegraphics[height=6cm]{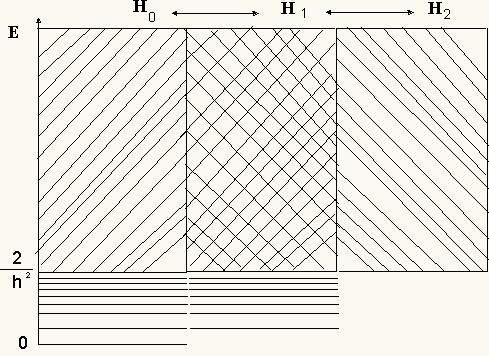}
\caption{The spectrum of the Kepler-Coulomb Hamiltonian (left panel). The spectrum of the supersymmetric
Kepler-Coulomb Hamiltonian (right panel). }
\end{figure}
\end{center}

\subsection{Spin-statistics structure of the supersymmetric Kepler/Coulomb problem}
 We have unveiled the spectrum of the planar supersymmetric Kepler/Coulomb problem solving the spectral problem of the two scalar Hamiltonians and using the supersymmetry algebra to obtain the eigenfunctions of the matrix operator.
 It is convenient, however, to look at the system as a whole, i.e., search directly for the spectrum of the $4\times 4$-matrix supersymmetric Hamiltonian operator.

 With this goal in mind we define the \lq\lq spin" operator
\[
\frac{1}{2}\hat{S} = -i\frac{\hbar}{2} \left(\hat{\psi}^\dagger_1\hat{\psi}_2-\hat{\psi}_2^\dagger\hat{\psi}_1\right)=-i\frac{\hbar}{2}\left(\begin{array}{cccc}
0 & 0 & 0 & 0 \\ 0 & 0 & 1 & 0 \\ 0 & -1 & 0 & 0 \\ 0 & 0 & 0 & 0 \end{array}\right) \quad .
\]
Clearly, $[\hat{S},\hat{N}]=0$, such that $\frac{1}{2}\hat{S}$ and $\hat{N}$ share eigenstates fulfilling a quantum mechanical spin-statistics theorem. The Bosonic eigenstates of $\hat{N}$ are zero spin eigenstates of $\frac{1}{2}\hat{S}$, whereas the Fermionic eigenstates of $\hat{N}$ are one-half spin eigenstates of $\frac{1}{2}\hat{S}$:
\[
\frac{1}{2}\hat{S}\left(\begin{array}{c} \psi^{(0)} \\ 0 \\ 0 \\ 0 \end{array}\right)=\frac{1}{2}\hat{S}\left(\begin{array}{c} 0 \\ 0 \\ 0 \\ \psi^{(2)}  \end{array}\right)=0 \qquad , \qquad \frac{1}{2}\hat{S}\left(\begin{array}{c} 0 \\ \psi^{(1)} \\ \pm i \psi^{(1)} \\ 0 \end{array}\right)=\frac{\hbar}{2}\left(\begin{array}{c} 0 \\ \psi^{(1)} \\ \pm i \psi^{(1)} \\ 0 \end{array}\right)\quad .
\]

Note also that neither the orbital angular momentum $\hat{L}$ nor the spin angular momentum $\hat{S}$
commute with $\hat{H}$. The \lq\lq total" angular momentum is the quantum invariant associated with simultaneous
rotations of the Bosonic $x_k$ and Fermionic $\psi_k$ coordinates{\footnote{Recall that $[\hat{L},x_k]=\hbar\varepsilon_{kj}x_j$ and $[\hat{S},\hat{\psi}_k]=\hbar\varepsilon_{kj}\hat{\psi}_j$,
$\varepsilon_{12}=-\varepsilon_{21}=1$ , $\varepsilon_{11}=\varepsilon_{22}=0$ .} }:
\[
\hat{J} = \hat{L}+\hat{S} =-i\hbar \left( x_1\frac{\partial}{\partial x_2}-x_2\frac{\partial}{\partial x_1}\right)-i\hbar\left(\hat{\psi}^\dagger_1\hat{\psi}_2-\hat{\psi}_2^\dagger\hat{\psi}_1 \right)\quad .
\]
Therefore, besides the fact that $[\hat{J},\hat{N}]=0$, one can use $[\hat{J},\hat{g}]=[\hat{J},\hat{X}]=0$ to show that:
\[
[\hat{J},\hat{Q}]=[\hat{J},\hat{Q}^\dagger ]=[\hat{J},\hat{H}]=0 \, \, \, ,
\]
where $\hat{X}$ is defined as in (\ref{hedg}):
\[
\hat{X}=[\hat{g},\hat{g}^\dagger]={\mathbb I}_4-2\hat{N}+2\hat{g}^\dagger\hat{g} \quad .
\]
We have two Clifford supersymmetric operators commuting with each other: ${\hat H}$ and ${\hat J}$. The supersymmetric system as a whole is integrable. Now, the challenge is to find more Clifford differential operators commuting with the Hamiltonian. In \cite{KLPW} the authors found the supersymmetric version of the Runge-Lenz vector operator -henceforth, the supersymmetric KLPW vector operator-:
\[
\hat{W}_1 =\frac{1}{2}\left(\hat{p}_2\hat{J}+\hat{J}\hat{p}_2\right)-\frac{ x_1}{r}\cdot\hat{X} \, \,
\quad , \quad \quad  \hat{W}_2 =-\frac{1}{2}\left(\hat{p}_1\hat{J}+\hat{J}\hat{p}_1\right)-\frac{ x_2}{r}\cdot\hat{X}\quad .
\]
One could guess the step from $\hat{L}$ to $\hat{J}$ and the need for the factor $\hat{X}$ is also no surprise
given its r$\hat{\rm o}$le in the supersymmetric Hamiltonian ${\hat H}$. A long computation ensures that the two components of this $4\times 4$-matrix vector differential operator will indeed commute with the Hamiltonian and with the Fermi number operator:
\[
[\hat{W}_k,\hat{Q}] = [\hat{W}_k,\hat{Q}^\dagger]=[\hat{W}_k,\hat{H}]=[\hat{W}_k,\hat{N}]=0 \quad .
\]
Some work is also necessary to check that
\[
[\hat{J},\hat{W}_1]=i\hbar \hat{W}_2 \quad , \quad [\hat{J},\hat{W}_2]=-i\hbar \hat{W}_1 \qquad , \qquad
[\hat{W}_1,\hat{W}_2]=-2 i \hbar \left(\hat{H}-\frac{2}{\hbar^2}\right)\hat{J} \qquad.
\]
Therefore, defining
\[
\hat{C}_k=\frac{\, \hat{W}_k}{\sqrt{2(\frac{2}{\hbar^2}-\hat{H})}}
\]
the ${\mathbb S}{\mathbb O}(3)$ Lie algebra is now closed -in the
sub-space of states of energy in the range
$E\in(0,\frac{2}{\hbar^2})$- by the Clifford operators $\hat{C}_1,
\hat{C}_2, \hat{C}_3=\hat{J}$
\[
 [\hat{C}_a,\hat{C}_b]=-i\hbar \varepsilon_{abc}\hat{C}_c
 \, \, \, ,  \quad a,b,c=1,2,3
 \]
 and the Casimir operator is the $4\times 4$-matrix differential operator: $
 \hat{{\cal C}}^2=\frac{1}{\hbar^2}\left[\hat{C}_1^2+\hat{C}^2_2+\hat{C}_3^2\right]$. In \cite{KLPW}
 the authors were able to find:
 \[
\hat{H}\left|_{\hat{Q}{\cal H}}\right.=\frac{1}{2}\hat{Q}\hat{Q}^\dagger =\frac{2}{\hbar^2}\left(1-\frac{(1-2\hat{N})^2}{(1-2\hat{N})^2+4\hat{\cal C}^2}\right)
  \, , \, \, \, \hat{H}\left|_{\hat{Q}^\dagger{\cal H}}\right.=\frac{1}{2}\hat{Q}^\dagger\hat{Q} =\frac{2}{\hbar^2}\left(1-\frac{(3-2\hat{N})^2}{(3-2\hat{N})^2+4\hat{\cal C}^2}\right)\quad ,
  \]
such that $E_j^{(0)}=\frac{2}{\hbar^2}\left(1-\frac{1}{(2j+1)^2}\right)=E^{(1)}_j$, the bound state energies
paired through supersymmetry in ${\cal S}{\cal H}_0$ and ${\cal S}{\cal H}_1$, reappear.

\section{The planar quantum Euler/Coulomb problem and supersymmetry}
Our second task is to build a supersymmetric quantum mechanical system inspired in the Euler/Coulomb problem:
a massive/charged particle that moves on a plane under the influence of two fixed Newtonian/Coulombian centers, see next Figure.

\begin{center}
\begin{figure}[ht]
\hspace{5cm}\includegraphics[height=3.5cm]{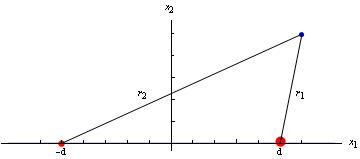} \caption{
Location of the two centers and distances to the particle from the
centers.}
\end{figure}
\end{center}

\subsection{The quantum Euler/Coulomb Hamiltonian}
The quantum Euler Hamiltonian is:

\[ \hat{I}_1= -\frac{\hbar^2 }{2 m} \bigtriangleup - \frac{\alpha_1}{r_1} - \frac{\alpha_2}{r_2} \quad .
\]
Here, $(x_1^\pm=\pm d,x_2^\pm=0)$ are the locations of the centers in the $x_1$-axis, and $\alpha_1=\alpha
\geq \alpha_2=\delta\alpha$ are the center strengths. With no loss of generality, we assume $\delta \in (0,1]$
 affording hetero-nuclear one-electron diatomic molecular ions. Thus, the strengths depend on the atomic numbers
 $Z_1$ and $Z_2$ of the atoms and the $\delta$ parameter is a positive rational number less than or equal to one:
\[
\alpha=e^2 Z_1\, \, \, \quad  , \quad \delta=\frac{Z_2}{Z_1} \leq 1 , \, \, , \quad Z_1,Z_2\in {\mathbb N}^* \, \, , \, \, Z_1\geq Z_2 \qquad .
\]
Finally, the distances of the particle to the two centers are: $r_1=\sqrt{(x_1-d)^2 + x_2^2}$
, $r_2=\sqrt{(x_1+d)^2+x_2^2}$.

Unlike in the Kepler/Coulomb problem there is a parameter with dimensions of length in the system: the distance between the centers $d$. This allows us to use non dimensional spatial coordinates:
\[
x_1\rightarrow  d\, x_1 \, \, , \, \,  x_2\rightarrow  d\, x_2
\quad , \quad
r_1\rightarrow d\, r_1=d\sqrt{(x_1-1)^2 + x_2^2} \, \, , \, \,
r_2\rightarrow d\, r_2=d\sqrt{(x_1+1)^2+x_2^2} \, \, \, .
\]
Note that there is also a fundamental action $\sqrt{m d\alpha}$ built from the parameters of the system
that provides a non dimensional Planck constant: $\bar{\hbar}=\frac{\hbar}{\sqrt{m d\alpha }}$.
Assembling all this together, the linear momentum and Hamiltonian operators go to $\hat{p}_i \, \rightarrow \, \sqrt{\frac{m \alpha}{d}}\hat{p}_i$ and $\hat{I}_1\, \rightarrow \, \frac{\alpha}{d}\hat{I}_1$,
where the new non dimensional operators are:
\[ \hat{p}_i=-i\bar{\hbar}\frac{\partial}{\partial x_i} \quad , \quad \hat{I}_1= -\frac{\bar{\hbar^2}}{2} \left(
\frac{\partial^2}{\partial x_1^2}+ \frac{\partial^2}{\partial
x_2^2}\right) - \frac{1}{r_1} - \frac{\delta}{r_2} \quad .
\]

In this problem there is a non-obvious symmetry operator $\hat{I}_2 \, \rightarrow \, m d \alpha \hat{I}_2$
where the non dimensional operator reads \cite{Perelomov}:
\[
\hat{I}_2=\frac{1}{2}\left(\hat{L}^2-\hat{p}_2^2\right)+\frac{1-x_1}{r_1}+\frac{(1+x_1)\delta}{r_2} \quad .
\]
 Just as the Runge-Lenz vector is quadratic in the momenta but unlike in the Kepler/Coulomb problem there are no more invariants in the Euler system which, accordingly, is only integrable. Explicitly,
\[ \hat{I}_2= -\frac{\bar{\hbar}^2}{2} \left( (x_1^2-1)
\frac{\partial^2}{\partial x_2^2} + x_2^2 \frac{\partial^2}{\partial
x_1^2} - 2 x_1 x_2 \frac{\partial^2}{\partial x_1 \partial x_2} -
x_1 \frac{\partial}{\partial x_1} - x_2 \frac{\partial}{\partial
x_2} \right)+\frac{1-x_1}{r_1}+\frac{(1+x_1)\delta}{r_2}
\]
and a little algebra shows that: $[\hat{I}_1, \hat{I}_2]=\hat{I}_1\hat{I}_2-\hat{I}_2\hat{I}_1=0$ .
\subsection{Separability of the Schr$\ddot{\rm o}$dinger equation in elliptic coordinates}

Because of the quadratic in the momenta symmetry operator, we expect that the Schr$\ddot{\rm o}$dinger equation
will be separable in some coordinate system on the plane. To skip the singularities in the centers one can
cover the plane by two open charts: the first chart is the open set ${\mathbb R}^+\times {\mathbb N}$ in
${\mathbb R}^2/ \{(-1,0)\}={\mathbb R}^+\times{\mathbb S}^1$ where ${\mathbb N}$ is the open North hemisphere
in ${\mathbb S}^1$. The second chart is the open set ${\mathbb R}^+\times {\mathbb S}$ in
${\mathbb R}^2/ \{(1,0)\}={\mathbb R}^+\times{\mathbb S}^1$ where ${\mathbb S}$ is the open South hemisphere.
Both charts must be glued at the abscissa axis $x_2=0$. Totally adapted to this topological situation are the
elliptic coordinates; the half-sum and half-difference of the distances to the centers of the particle:
\[
u=\frac{1}{2}(r_1+r_2)\in (1,+\infty)\hspace{1.5cm} ,
\hspace{1.5cm} v=\frac{1}{2} (r_2-r_1)\in (-1,1)
\]
that parametrize a two dimensional infinite strip ${\mathbb E}^2$. The Cartesian coordinates are obtained through the change
\[
x_1=u v \in (-\infty,+\infty)
\hspace{1.5cm} , \hspace{1.5cm} x_2=\pm \sqrt{(u^2-1) (1-v^2)} \in
(-\infty,+\infty)\quad ,
\]
which is a two-to-one map -one per chart- from ${\mathbb R}^2$ to ${\mathbb E}^2$ except at the $x_2$-axis, which is one-to-one mapped at the boundary of ${\mathbb E}^2$: $\partial{\mathbb E}^2= \{ (u=1, v), (u,v=\pm 1)\}$.

The Euler Hamiltonian in elliptic coordinates is of the separable form
\[
\hat{I}_1 = \frac{1}{u^2-v^2} \left( \hat{H}_u + \hat{H}_v \right)
\]
\[
\hat{H}_u = -\frac{\bar\hbar^2}{2} \left( (u^2-1)
\frac{\partial^2}{\partial u^2} + u \frac{\partial}{\partial u}
\right)  - (1+\delta) u \quad , \quad
\hat{H}_v= - \frac{\bar\hbar^2}{2} \left( (1-v^2)
\frac{\partial^2}{\partial v^2} - v \frac{\partial}{\partial v}
\right)  - (1-\delta) v
\]
and the symmetry operator also separates:
\[
\hat{I}_2 = \frac{1}{u^2-v^2} \left[ (u^2-1) \hat{H}_v - (1-v^2)
\hat{H}_u \right] \qquad .
\]
The ansatz $\psi_E(u,v) = \eta_E (u) \xi_E(v)$ converts the spectral problem
\begin{equation}
\hat{I}_1 \psi_E(u,v) = E \psi_E(u,v) \label{scr2c}
\end{equation}
into separable:
\begin{eqnarray} && -\bar{\hbar}^2(u^2-1)
\frac{d^2\eta_E}{d u^2}(u)-\bar{\hbar}^2 u \frac{d\eta_E}{d u}(u)- \left[ 2(1+\delta) u  + 2 u^2 E
 \right] \eta_E(u)= I  \eta_E(u) \label{scr2c1}
\\&&
-\bar{\hbar}^2(1-v^2) \frac{d^2\xi_E}{d u^2}(v)+\bar{\hbar}^2 v \frac{d \xi_E}{d u}(v)+ \left[ -
2(1-\delta) v  + 2 v^2 E  \right] \xi_E(v)=- I \xi_E(v) \label{scr2c2}\qquad .
\end{eqnarray}
The Sch$\ddot{\rm o}$dinger PDE equation (\ref{scr2c}) becomes the
two coupled ODE's (\ref{scr2c1})-(\ref{scr2c2}) where the separation
constant $I$ is the eigenvalue of the symmetry operator $ \hat{I} =
- 2 \hat{I}_1 - 2 \hat{I}_2 $.

 We could try to solve (\ref{scr2c1})-(\ref{scr2c2}) directly but we still perform the following
 change of variables:
 \begin{equation}
 x=\frac{1}{2}{\rm arccosh}u \in [0,\infty) \qquad , \qquad y=\frac{1}{2}{\rm arccos}v\in
[0,\frac{\pi}{2}] \label{schco} \qquad .
 \end{equation}
 Equation  (\ref{scr2c1}) becomes the Razavy equation (\ref{scre}) \cite{Razavy1}, and (\ref{scr2c2})
 becomes the Razavy trigonometric (\ref{scrte}) or Whittaker-Hill equation \cite{Razavy2}:
 \begin{eqnarray}  -{\displaystyle
\frac{d^2\eta_E(x)}{dx^2}}+ \left( \zeta \, \cosh 2x- M \right) ^2
\eta_E(x)&=&  \lambda \eta_E(x)  \label{scre}
\\  {\displaystyle \frac{ d^2 \xi_E(y)}{d y^2}} + ( \beta \, {\rm cos} 2 y - N)^2 \xi_E(y)
&=& \mu \xi_E(y) \label{scrte} \quad .
\end{eqnarray}
The parameters in the Razavy equations (\ref{scre}) and (\ref{scrte}) are defined in terms of the energy
and the eigenvalue of the symmetry operator in the form:
\begin{eqnarray*}
\zeta &=& \frac{2 \sqrt{ - 2 E}}{\bar{\hbar}} \quad , \quad M^2 =-\frac{2 (1+\delta)^2}{\bar{\hbar}^2 E} \quad , \quad \lambda=
M^2 + \frac{4 I}{\bar{\hbar}^2}
 \\
\beta &=& - \frac{2 \sqrt{-2 E}}{\bar{\hbar} } \quad , \quad N^2= -\frac{2
(1-\delta)^2}{\bar{\hbar}^2 E} \quad , \quad
\mu= N^2 + \frac{4 I}{\bar{\hbar}^2}  \qquad .
\end{eqnarray*}
We stress the following subtle point: the Razavy equations are defined for fixed $\zeta$, $M$, and $\lambda$ or
for fixed $\beta$, $N$, and $\mu$. We obtain, however, Razavy equations for parameters determined from $E$ and $I$.
Therefore, we address an infinite number of entangled Razavy and Razavy trigonometric equations.

In Reference \cite{FGR} it is shown that the Razavy and Whittaker-Hill equations for $M$ and $N$ positive integers are quasi-exactly solvable (QES) systems and all the finite solutions -polynomials times fast decreasing exponentials- are found by algebraic means. Our strategy will be to use this information in the search for the bound state eigenvalues and eigenfunctions of the Euler/Pauli Hamiltonian. Our results have been partially published in \cite{MAJM}. Thus, we shall not repeat the analysis here. Instead, we shall develop the program in the supersymmetric version of the Euler problem.

\subsection{The supersymmetric modified Euler/Coulomb Hamiltonian}

In \cite{MAJM} we gave arguments for selecting the following superpotential
\begin{equation}
W(x_1, x_2)=-\frac{2}{\bar\hbar}\left(r_1+\delta r_2\right) \label{2csup}
\end{equation}
in order to develop a supersymmetric quantum mechanical system from two fixed centers containing
a mild deformation of the Euler/Coulomb system in the $\hat{N}=0$ sector. In fact, from the superpotential partial derivatives
\begin{eqnarray*}
&& \frac{\partial W}{\partial x_1}= -\frac{1}{2\bar\hbar}\left(\frac{x_1-1}{r_1}(1-\delta)+\frac{x_1+1}{r_2}(1+\delta)\right)
\quad , \quad \frac{\partial W}{\partial x_2}= -\frac{1}{2\bar\hbar}\left(\frac{x_2}{r_1}(1-\delta)+\frac{x_2}{r_2}(1+\delta)\right) \\ && \frac{\partial^2 W}{\partial x_1^2}=  -\frac{1}{2\bar\hbar}\left\{\left(\frac{1}{r_1}-\frac{(x_1-1)^2}{4 r_1^3}\right)(1-\delta)+\left(\frac{1}{r_2}-\frac{(x_1+1)^2}{4 r_2^3}\right)(1+\delta)\right\}
 \\ && \frac{\partial^2 W}{\partial x_2^2}= -\frac{1}{2\bar\hbar}\left\{\left(\frac{1}{r_1}-\frac{x_2^2}{4 r_1^3}\right)(1-\delta)+\left(\frac{1}{r_2}- \frac{x_2^2}{4 r_2^2}\right)(1+\delta)\right\}\\ && \frac{\partial^2 W}{\partial x_1\partial x_2} =\frac{1}{2\bar\hbar}\left(\frac{x_2(x_1-1}{4 r_1^3}(1-\delta)+\frac{x_2(x_1+1)}{4r_2^3}\right)= \frac{\partial^2 W}{\partial x_2\partial x_1}
\end{eqnarray*}
we obtain first the supercharges. In turn, the scalar Hamiltonians,
\begin{eqnarray*}
\hat{H}_0&=& -\frac{\bar\hbar^2}{2}\bigtriangleup+\frac{2}{\bar\hbar^2}\left[1+\delta^2+\delta\left(\frac{r_1}{r_2}+
\frac{r_2}{r_1}-\frac{4}{r_1r_2}\right)\right]-\frac{1}{r_1}-\frac{\delta}{r_2}\\ \hat{H}_2&=& -\frac{\bar\hbar^2}{2}\bigtriangleup+\frac{2}{\bar\hbar^2}\left[1+\delta^2+\delta\left(\frac{r_1}{r_2}+
\frac{r_2}{r_1}-\frac{4}{r_1r_2}\right)\right]+\frac{1}{r_1}+\frac{\delta}{r_2}\quad ,
\end{eqnarray*}
and the matrix Hamiltonian:
{\small\[
\hat{H}_{1}=\left(\begin{array}{cc}\frac{1}{2}\left( \hat{H}_0+\hat{H}_2\right)-
\frac{(x_1-1)^2-x^2_2}{r_1^3} -
\delta\frac{(x_1+1)^2-x_2^2}{r_2^3}  &
-2  \left(
\frac{x_2(x_1-1)}{r_1^3} + \delta\frac{x_2(x_1+1)}{r_2^3} \right)
\\  -2  \left(
\frac{x_2(x_1-1)}{r_1^3} + \delta\frac{x_2(x_1+1)}{r_2^3} \right) & \frac{1}{2}\left( \hat{H}_0+\hat{H}_2\right) +
\frac{(x_1-1)^2-x^2_2}{r_1^3} +
\delta\frac{(x_1+1)^2-x_2^2}{r_2^3}  \end{array}\right) \,
\]}
are derived. Now, the rationale for the choice of the superpotential (\ref{2csup}) is clearer: at the limit where the two centers are superposed, $d=0$ and $r_1=r_2$, the superpotential, the supercharges and the superHamiltonian become those of the supersymmetric Kepler/Coulomb problem (with non-dimensional strength $1+\delta$ instead of $1$).

In this one-parametric deformation of the Kepler problem a very subtle conundrum arises. In the Kepler/Coulomb case, the non-supersymmetric $\hat{K}$ and the $\hat{N}=0$ scalar $\hat{H}_0$ Hamiltonians only differ in the shift by a constant necessary to push the ground state energy to zero as required by supersymmetry. In the Euler/Coulomb case, $\hat{I}_1$ and $\hat{H}_0$ differ in a non-constant potential such that in the Kepler limit becomes the constant shift{\footnote{We temporarily come back to dimensional coordinates in order to see the limit.}}:
\[
V_S(x_1,x_2)=\frac{2m\alpha^2}{\hbar^2}\left[1+\delta^2+\delta\left(\frac{r_1}{r_2}+
\frac{r_2}{r_1}-\frac{4 d^2}{r_1r_2}\right)\right] \ , \ \lim_{d\to 0}\,
V_S(x_1,x_2)=\frac{2 m \alpha^2 (1+\delta)^2}{\hbar^2} \ \, .
\]
The r$\hat{\rm o}$le of this potential is to shift the negative bound state energies in a harmless way: $\hat{H}_0$, like $\hat{I}_1$, still admits a symmetry operator that is quadratic in the momenta  and the supersymmetric spectral problem in ${\cal S}{\cal H}_0$ is still separable in elliptic coordinates !! In other words, we choose the superpotential in such a way that the separability in elliptic coordinates of $\hat{H}_0$ is preserved even though we must add a \lq\lq classical" piece -important when $\bar\hbar$ tends to $0$ - to the Euler/Coulomb non-supersymmetric Hamiltonian.

In fact, $\hat{H}_0$ and the symmetry operator $\hat{I}_2$ in elliptic coordinates are still of the form
\[
\hat{H}_0=\frac{1}{u^2-v^2}\left(\hat{H}_u^{(0)}+\hat{H}_v^{(0)}\right) \qquad , \qquad \hat{I}_2^{(0)} = \frac{1}{u^2-v^2} \left[ (u^2-1) \hat{H}_v^{(0)} - (1-v^2)
\hat{H}_u^{(0)} \right]
\]
where $\hat{H}_u^{(0)}$ and $\hat{H}_v^{(0)}$ are now {\footnote{Obviously, the same situation happens in the $\hat{N}=2$ sector. $\hat{H}_2$ and $\hat{I}_2^{(2)}$ are given in the same way in terms of $\hat{H}_u^{(2)}$ and $\hat{H}_v^{(2)}$ that differ from $\hat{H}_u^{(0)}$ and $\hat{H}_v^{(0)}$ in the sign of the last terms.}}:
\begin{eqnarray} \hat{H}_u^{(0)} &=& -\frac{\bar\hbar^2}{2}\left((u^2-1)\frac{\partial^2}{\partial
u^2}+u\frac{\partial}{\partial u}\right)+\frac{2}{\bar\hbar^2}(1+\delta)^2 (u^2-1)-(1+\delta)u \label{at2ceu}\\ \hat{H}_v^{(0)} &=&
-\frac{\bar\hbar^2}{2}\left((1-v^2)\frac{\partial^2}{\partial
v^2}- v\frac{\partial}{\partial v}\right)+ \frac{2}{\bar\hbar^2}(1-\delta)^2 (1-v^2)-(1-\delta)v \label{at2cev}\quad .
\end{eqnarray}

The separation ansatz $\psi^{(0)}_E(u,v)=\eta^{(0)}_E(u)\xi^{(0)}_E(v)$ plugged into
the supersymmetric spectral problem in the $\hat{N}=0$ Bosonic subspace
\[
\hat{H}_0\psi^{(0)}_E(u,v)=E\psi^{(0)}_E(u,v)
\]
reduces the PDE Schr$\ddot{\rm o}$dinger equation to the system of separated ODE's
\begin{eqnarray} \hspace{-1.3cm}&&
\left[-\bar\hbar^2(u^2-1)\frac{d^2}{du^2}-\bar\hbar^2u\frac{d}{du}+
\left(4 \frac{(1+\delta)^2}{\bar{\hbar}^2} (u^2-1) - 2 (1+\delta)
u - 2 E
 u^2 \right)
\right] \eta^{(0)}_E(u)=I \eta^{(0)}_E(u) \label{ss2ce1}
\\ \hspace{-1.3cm} &&
\left[-\bar\hbar^2(1-v^2)\frac{d^2}{dv^2}+\bar\hbar^2v\frac{d}{dv}+
\left(4 \frac{(1-\delta)^2}{\bar{\hbar}^2} (1-v^2) - 2 (1-\delta)
v +2 E v^2 \right) \right] \xi^{(0)}_E(v) = - I \xi^{(0)}_E(v)\label{ss2ce2}
\, \, ,
\end{eqnarray}
where the separation constant $I$ is the eigenvalue of the symmetry
operator $\hat{I}=-2 \hat{H}_0-2 \hat{I}_2^{(0)}$.

\subsection{Bound states from entangled Razavy and Whittaker-Hill equations}

 The change of coordinates (\ref{schco}) transforms (\ref{ss2ce1}) and (\ref{ss2ce2}) respectively in the
 Razavy and Whittaker-Hill (three-term Hill) equations
\begin{eqnarray} -\frac{d^2\eta_E^{(0)}(x)}{dx^2}+ \left(
\zeta \cosh 2x- M \right) ^2 \eta_E^{(0)}(x)&=& \lambda\,
\eta_E^{(0)}(x)  \label{uveq1}\\   \frac{ d^2 \xi_E^{(0)}(y)}{ d
y^2} + ( \beta {\rm cos} 2 y - N)^2 \, \xi_E^{(0)}(y) &=&
\mu\ \xi_E^{(0)}(y) \label{uveq2} \qquad ,
\end{eqnarray}
where the parameters are now:
\begin{eqnarray*}
 && \zeta=  \frac{2}{\bar{\hbar}} \sqrt{\frac{4}{\bar{\hbar}^2} (1+\delta)^2 - 2
 E} \quad , \quad M^2
=\frac{2 (1+\delta)^2}{ 2 (1+\delta)^2 -\bar\hbar^2 E} \quad , \quad  \lambda= M^2 + \frac{4}{\bar{\hbar}^2}
(I + \frac{4(1+\delta)^2}{\bar\hbar^2})
 \\ &&
\beta = - \frac{2}{\bar{\hbar}} \sqrt{\frac{4}{\bar\hbar^2}
(1-\delta)^2 - 2 E} \quad , \quad N^2 = \frac{2 (1-\delta)^2 }{ 2
(1-\delta)^2 - \bar{\hbar}^2 E} \quad , \quad \mu = N^2 +
\frac{4}{\bar{\hbar}^2} (I + \frac{4 (1-\delta)^2}{\bar{\hbar}^2} )
\qquad .
\end{eqnarray*}
\subsection{Eigenfunctions from the Razavy equations}

In \cite{FGR} it is shown that the Razavy equation (\ref{uveq1}) is a quasi-exactly solvable algebraic equation that admits $n+1$ finite (polynomial times decaying exponential) solutions if $ M = n + 1$ and $n \in \mathbb{N}$ is a natural number. Moreover, there are $n+1$ solutions for $n+1$ different values of $\lambda$ characterized by an integer $m=1,2,3, \cdots , n+1$. All the eigenvalues between $0$ and $\frac{2(1+\delta)^2}{\bar\hbar^2}$ of the supersymmetric modified Euler/Coulomb spectral problem in ${\cal S}{\cal H}_0$ are obtained in this way:
\[
M=n+1 \, \, \, \Rightarrow \, \, \, \, E_n^{(0)}=\frac{2(1+\delta)^2}{\bar\hbar^2}\left(1-\frac{1}{(n+1)^2}\right)\qquad .
\]
Concerning the eigenfunctions, we search for solutions of the $M=n+1$ Razavy equations by means of the series expansion {\footnote{This ansatz, and others related to this, allows to represent the Razavy Hamiltonians in terms of differential operators that belong to the enveloping algebra of ${\mathbb S}{\mathbb L}(2,{\mathbb R})$, see \cite{FGR} and \cite{Finkel}.}}:
\[
 \eta_{n}(z) = z^{-\frac{n}{2}} \, e^{-\frac{\zeta_n}{4}
\left( z + \frac{1}{z}\right)} \sum_{k=0}^\infty \frac{(-1)^k \,
P_k(\lambda)}{(2\zeta)^k\,k!}\, z^k \quad , \quad z=e^{2x} \quad , \quad \zeta_n=\frac{4(1+\delta)}{\bar\hbar^2(n+1)}
\]
where $P_k(\lambda)$ are polynomials of order $k$ in $\lambda$ to be fixed. The ODE Razavy equation is then solved if the following three-term recurrence relations among the polynomials hold:
\[
P_{k+1}(\lambda)=\left( \lambda - ( 4 k (n-k) + 2 n +1 +
\zeta_n^2) \right) \, P_k(\lambda)-( 4 k ( n +1-k) \zeta_n^2) \,
P_{k-1}(\lambda)\ ,\ k\geq 0\quad .
\]
In particular, if $\lambda_{n m}$, $m=1,2, \cdots , n+1$, is one of the  $n+1$ roots of $P_{n+1}$,
$P_{n+1}(\lambda_{n m})=0$, then
\[
0=P_{n+2}(\lambda_{n m}) = P_{n+3}(\lambda_{nm})=P_{n+4}(\lambda_{nm})=\dots \dots
\]
and the series  truncates. The degeneracy in the energy is broken by the eigenvalues $I$ of the
symmetry operator $\hat{I}_2^{(0)}$ provided by the roots $\lambda_{nm}$ in the form
\[
I_{n m} = \frac{\bar{\hbar}^2}{4} ( \lambda_{n m}
- (n+1)^2) - \frac{4(1+\delta)^2}{\bar{\hbar}^2} \quad ,
\]
which distinguishes between the different polynomials $P_m(\lambda_{nm})$.

We solve the finite-step recurrences for the lower-energy cases
\begin{itemize}

\item {\bf $n=0$}:
\[
\zeta_0=\frac{4(1+\delta)}{\bar\hbar^2}\, \, , \, \, P_0(\lambda)=1 \, \, , \, \, P_1(\lambda)=\lambda-(1+\zeta_0^2) \quad , \quad \lambda_{01} = 1 + \zeta_0^2
\]

\item {\bf $n=1$}:
\begin{eqnarray*}
\zeta_1&=&\frac{2(1+\delta)}{\bar\hbar^2}\quad , \quad P_0(\lambda)=1 \quad , \quad P_1(\lambda)=\lambda-(3+\zeta_1^2) \\ P_2(\lambda)&=&\lambda^2-2(3+\zeta_1^2)\lambda+9+2\zeta_1^2+\zeta_1^4
\\ \lambda_{11} &=& 3 - 2 \zeta_1 +
\zeta_1^2 \qquad \quad , \qquad \quad \lambda_{12} = 3 + 2 \zeta_1 +
\zeta_1^2
\end{eqnarray*}

\item {\bf $n=2$}

 \begin{eqnarray*}
 \zeta_2&=&\frac{4(1+\delta)}{3\bar\hbar^2} \quad , \quad P_0(\lambda)=1 \quad , \quad P_1(\lambda)=\lambda-(5+\zeta_2^2) \\ P_2(\lambda)&=& (\lambda-(5+\zeta_2^2))(\lambda-(9+\zeta_2^2))-8\zeta_2^2
  \\ P_3(\lambda)&=&(\lambda-(5+\zeta_2^2))\left[(\lambda-(5+\zeta_2^2))(\lambda-(9+\zeta_2^2))-16\zeta_2^2\right]\\ \lambda_{21}&=& \zeta_2^2 + 5 \quad,\quad \lambda_ {22}=\zeta_2^2 + 7 - 2
\sqrt{1 + 4\zeta_2^2 } \quad,\quad \lambda_ {23}=\zeta_2^2 + 7 + 2
\sqrt{ 1 + 4\zeta_2^2 }
\end{eqnarray*}
\end{itemize}
and show the results for the lower eigenvalues and eigenfunctions in the next Table:
\begin{table}[htdp]
\begin{center}
\begin{tabular}{|c|c|c|} \hline  Energy & Eigen-function & Separation constant \\ \hline & & \\{\tiny $E_0^{(0)}=0$} & {\tiny $\eta_{01}^{(0)}(u) = e^{- \frac{2 (1+\delta)u}{\bar{\hbar}^2}}$ } & {\tiny $I_{01}^{(0)} = 0 $}\\ \hline & & \\ {\tiny $E_1^{(0)}=\frac{3(1+\delta)^2}{2\bar\hbar^2}$} & {\tiny $\left\{
\begin{array}{c}
 \eta_{11}^{(0)} (u) = e^{- \frac{(1+\delta)u}{\bar{\hbar}^2}}
\sqrt{2 (u+1)}  \\ \\
\eta_{12}^{(0)}(u) =- e^{- \frac{(1+\delta)u}{\bar{\hbar}^2}}
\sqrt{2(u-1)}
\end{array}
\right.$} & {\tiny $\left\{
\begin{array}{c}
 I_{11}^{(0)} =  -\frac{\bar{\hbar}^2}{4} - \frac{3(1+\delta)^2}{\bar{\hbar}^2} -
 (1+\delta)  \\  \\
I_{12}^{(0)} = -\frac{\bar{\hbar}^2}{4} - \frac{3
(1+\delta)^2}{\bar\hbar^2} +(1+\delta)
\end{array}
\right.$}
\\ \hline & & \\{\tiny $E_2^{(0)}=\frac{16(1+\delta)^2}{
9\bar\hbar^2}$} & {\tiny $\left\{\begin{array}{c}\eta_{21}^{(0)}(u) = - 2 e^{-\frac{2(1+\delta)u}{3\bar{\hbar}^2}} \sqrt{u^2 - 1}  \\ \\
\eta_{22}^{(0)}(u) = \frac{3 \bar{\hbar}^2}{4(1+\delta)} e^{-
\frac{2(1+\delta)u}{ 3 \bar{\hbar}^2}} \left[ \frac{8 (1+\delta) }{3
\bar{\hbar}^2} u - 1 + \sqrt{ 1 + \frac{ 64 (1+\delta)^2}{ 9
\bar{\hbar}^4}} \right] \nonumber \\ \\
\eta_{23}^{(0)}(u) = \frac{3 \bar{\hbar}^2}{4(1+\delta)} e^{-
\frac{2 (1+\delta) u}{3 \bar{\hbar}^2}} \left[ \frac{8 (1+\delta)
}{3 \bar{\hbar}^2} u - 1 - \sqrt{ 1 + \frac{ 64 (1+\delta)^2}{ 9
\bar{\hbar}^4}} \right]\nonumber
\end{array}
\right.$} & {\tiny $ \left\{ \begin{array}{c} I_{21}^{(0)} = - \bar{\hbar}^2 - \frac{32  (1+\delta)^2}{ 9 \bar{\hbar}^2} \\ \\
I_{22}^{(0)} = -\frac{\bar{\hbar}^2}{2} - \frac{32 (1+\delta)^2}{9
\bar{\hbar}^2} - \frac{1}{6} \sqrt{ 64(1+\delta)^2 + 9
\bar{\hbar}^4}  \\ \\
I_{23}^{(0)} =  -\frac{\bar{\hbar}^2}{2} - \frac{32(1+\delta)^2}{9
\bar{\hbar}^2} + \frac{1}{6} \sqrt{64(1+\delta)^2 + 9 \bar{\hbar}^4}
\end{array}
\right.$}  \\
\hline
\end{tabular}
\end{center}
\end{table}

\subsubsection{Contribution from the Whittaker-Hill equations}

For these values of the eigenvalues $E_n$ and $I_{n m}$ two of the parameters of the Whittaker-Hill equations (\ref{uveq2}) become:
\[
\beta_n=-\frac{4}{\bar\hbar^2}\sqrt{\frac{(1+\delta)^2}{(n+1)^2}-4
\delta} \qquad , \qquad
\mu_{nm}=N_n^2+\frac{4}{\bar\hbar^2}\left(I_{nm}+\frac{4(1-\delta)^2}{\hbar^2}\right)
\]
but the other one $N_n=\frac{1}{\sqrt{1-\left(\frac{1+\delta}{1-\delta}\right)^2\left(1-\frac{1}{(n+1)^2}\right)}}$
is not a positive integer if $n\geq 1$ and $\delta=\frac{Z_2}{Z_1}$. The existence of solutions
of the equation
\[
Z_1\left[(k+1)\sqrt{n(n+2)}-(n+1)\sqrt{k(k+2)}\right]=Z_2\left[(k+1)\sqrt{n(n+2)}+(n+1)\sqrt{k(k+2)}\right]
\]
over positive integers $n$ and $k$ would be necessary to simultaneously find $M=n+1$ and $N=k+1$. It is clear that this is not the case and there are no finite solutions other than the ground state in the supersymmetric two-center problem. The analogous equation in the non supersymmetric case is:
\[
Z_1=\frac{n+k+2}{n-k}.Z_2 \quad ,
\]
which admits solutions for positive integers $n$ and $k$ found by Demkov and his colleagues over forty
years ago, see \cite{Demkov} and, e.g., \cite{Abramov}.

If $n=0$, $E_0^{(0)}=0$, $I_{01}^{(0)}=0$, however, the Whittaker-Hill equation is also QES, see \cite{FGR1}, with a unique finite wave function:
\[
\beta_{0}=-\frac{4}{\bar\hbar^2}(1-\delta) \quad , \quad N_{0}=1 \quad , \quad \mu_{01}=1+\frac{16(1-\delta)^2}{\bar\hbar^4} \quad , \quad  \xi_{01}^{(0)}(v)=e^{\frac{2(1-\delta)}{\bar\hbar^2}v} \quad .
\]
The analytic wave function of  the ground state in the ${\cal S}{\cal H}_0$ sector is:
\begin{equation}
E_0^{(0)}=0 \hspace{1cm} , \hspace{1cm} \psi_{01}^{(0)}(u,v)=\eta_{01}^{(0)}(u)\xi_{01}^{(0)}(v)={\rm exp}[-
\frac{2(1+\delta)u}{\bar\hbar^2}]\cdot{\rm
exp}[\frac{2(1-\delta)v}{\bar\hbar^2}] \quad ; \label{bzm2}
\end{equation}
thus, there is a normalizable Bosonic zero energy {\bf ground state}, a zero mode $\psi_{01}^{(0)}(u,v)$,
in the supersymmetric two-center system. Supersymmetry is not spontaneously broken.

In the WH equation for the above parameters we could try a solution of the form \cite{FGR1}:
\[
 \xi_{n}(w) = w^{\frac{1-N_n}{2}} \, e^{-\frac{\beta_n}{4}
\left( w + \frac{1}{w}\right)} \sum_{k=0}^\infty \frac{(-1)^k \,
Q_k(\mu_{nm})}{(2\beta_n)^k\,k!}\, w^k \qquad , \qquad w=e^{2iy} \quad ,
\]
which solves (\ref{uveq2}) if the \lq\lq recurrence" relations between the $Q$-polynomials hold:
\[
Q_{k+1}(\mu_n)=\left( \mu_n - ( 4 k (N_n-1-k) + 2 N_n -1 + \beta_n^2)
\right) \, Q_k(\mu_n)-( 4 k ( N_n-k) \beta_n^2) \, Q_{k-1}(\mu_n)\ ,\
k\geq 0\ .
\]
This strategy, however, is not useful in this situation because the WH equations are not QES if $n\geq1$ ($N\neq k+1$).

Instead, we consider the WH equations in their algebraic form:
\begin{eqnarray*}
&& \frac{d^2\xi_{nm}^{(0)}}{d v^2}-\frac{v}{1-v^2}\frac{d\xi^{(0)}_{nm}}{dv}+\frac{A_{nm}+B_{nm}v+C_{nm}v^2}{1-v^2}\xi_{nm}^{(0)}(v)=0 \\
&& A_{nm}=-\frac{I_{nm}}{\bar\hbar^2}-\frac{4(1-\delta)^2}{\bar\hbar^4} \quad , \quad B_{nm}=\frac{2(1-\delta)}{\bar\hbar^2} \quad , \quad C_{nm}=-\frac{2}{\bar\hbar^2}\left(E_n-\frac{2(1-\delta)^2
}{\bar\hbar^2}\right) \qquad .
\end{eqnarray*}
Now, following the standard theory of Hill equations, see \cite{Hoc}, we search for power series solutions:  if $k\in{\mathbb N}$ is a natural number,
\begin{eqnarray}
\xi_{nm}^{(0)}(v)&=&\sum_{k=0}^\infty\, c_{nm}^{(k)}\, v^k  \quad , \quad c_{nm}^{(2)}= -\frac{A_{nm}}{2}c_{nm}^{(0)} \quad , \quad c_{nm}^{(3)}=-\frac{1}{6}\left((A_{nm}-1)c_{nm}^{(1)}+B_{nm}c_{nm}^{(0)}\right)\nonumber\\ c_{nm}^{(k)}&=&-\frac{1}{k(k-1)}\left\{\left(A_{nm}-(k-2)^2\right)c^{(k-2)}_{nm}+B_{nm}c_{nm}^{(k-3)}+C_{nm}c_{nm}^{(k-4)}\right\}
\quad , \quad k\geq 4 \label{recr}
\qquad \quad .
\end{eqnarray}
We encounter fourth-term recurrence relations, a difficult situation to deal with, although the two basic solutions are easily characterized:
\begin{enumerate}
\item $c_{nm+}^{(0)}=1$, $c_{nm+}^{(1)}=0$: $\xi^{(0)}_{nm+}(v)=1+\sum_{k=2}^\infty\, c_{nm+}^{(k)}\, v^k$.

\item $c_{nm-}^{(0)}=0$, $c_{nm-}^{(1)}=1$: $\xi^{(0)}_{nm-}(v)=v+\sum_{k=2}^\infty\, c_{nm-}^{(k)}\, v^k$.
\end{enumerate}
These series converge in the open $(-1,1)$ interval and are extended to cover the singularities $v=\pm 1$
setting the values:
\begin{eqnarray}
\xi_{nm+}^{(0)}(1)=1+\sum_{k=2}^\infty\, c_{nm+}^{(k)} \qquad &,& \qquad \xi_{nm+}^{(0)}(-1)=1+\sum_{k=2}^\infty \, (-1)^k \, c_{nm+}^{(k)} \label{bvsol}\\ \xi_{nm-}^{(0)}(1)=1+\sum_{k=2}^\infty\, c_{nm-}^{(k)} \qquad &,& \qquad \xi_{nm-}^{(0)}(-1)=-1+\sum_{k=2}^\infty \, (-1)^k \, c_{nm-}^{(k)} \qquad . \nonumber
\end{eqnarray}
The series for the ground state, for instance, are easy to find:
\begin{eqnarray*}
&& \xi_{01}^{(0)}(v)={\rm exp}\left[ \frac{2(1-\delta)}{\bar\hbar^2}v\right] =\sum_{k=0}^\infty\, c^{(k)}_{01}v^k \qquad , \qquad c^{(k)}_{01}=
\frac{1}{k!}\left(\frac{2(1-\delta)}{\bar\hbar^2}\right)^k \quad , \quad c_{01}^{(0)}=\xi_{01}^{(0)}(0)=1\\ && \frac{d\xi^{(0)}_{01}}{dv}(v)=\frac{2(1-\delta)}{\bar\hbar^2}{\rm exp}\left[ \frac{2(1-\delta)}{\bar\hbar^2}v\right] =\sum_{k=1}^\infty\, k c^{(k)}_{01}v^{k-1} \quad , \quad c_{01}^{(1)}=\frac{d\xi_{01}^{(0)}}{dv}(0)=\frac{2(1-\delta)}{\bar\hbar^2}\\ && \xi_{01}^{(0)}(\pm 1)=\sum_{k=0}^\infty \,
\frac{(\pm 1)^k}{k!}\left(\frac{2(1-\delta)}{\bar\hbar^2}\right)^k ={\rm exp}\left[ \pm \frac{2(1-\delta)}{\bar\hbar^2}\right] \quad \qquad .
\end{eqnarray*}
Any other solution of the WH equations is obtained by specific linear combinations of the two basic solutions. We will choose linear combinations of the general form
\begin{equation}
\xi_{nm}^{(0)}(v)=c_+\xi_{nm+}^{(0)}(v)+c_-\xi_{nm-}^{(0)}(v) \label{vsol}
\end{equation}
In fact, any choice of $(c_+,c_-)\in{\mathbb C}^2$ in (\ref{vsol}) fixes the extension to the boundary of the elliptic strip ${\mathbb E}=[1,+\infty)\times [-1,1]$ of the wave functions $\psi_{nm}^{(0)}(u,v)=\eta_{nm}^{(0)}(u)\xi_{nm}^{(0)}(v)$. Id est, extensions of $\hat{H}_0$ in $L^2({\mathbb E})$ are determined from the values of $\psi_{nm}^{(0)}(u,v)$ at the boundary: $\psi_{nm}^{(0)}(1,v)$ and $\psi_{nm}^{(0)}(u,\pm 1)$. We remark that these extensions are not essentially self-adjoint in general; different eigenfunctions have a small overlap for generic values of $\bar\hbar$.

\subsection{Two centers of the same strength}
In the case $\delta=1$ when the strength of the centers is identical the supersymmetric spectral problem
in ${\cal S}{\cal H}_0$ simplifies remarkably:
\begin{eqnarray} &&
\left[-\bar\hbar^2(u^2-1)\frac{d^2}{du^2}-\bar\hbar^2u\frac{d}{du}+
\left( \frac{16}{\bar{\hbar}^2} (u^2-1) - 4 u - 2 E
 u^2 \right)
\right] \eta^{(0)}_E(u)=I \eta^{(0)}_E(u)\quad
\qquad  \label{eq:spec21}
\\&&
\left[-\bar\hbar^2(1-v^2)\frac{d^2}{dv^2}+\bar\hbar^2v\frac{d}{dv}+
2 E v^2 \right] \xi^{(0)}_E(v) = - I
\xi^{(0)}_E(v) \quad \qquad  .\label{eq:spec31}
\end{eqnarray}
Again, (\ref{eq:spec21}) is equivalent to a Razavy  equation
\begin{equation} -\frac{d^2\eta_E^{(0)}(x)}{dx^2}+ \left(
\zeta \cosh 2x- M \right) ^2 \eta_E^{(0)}(x)= \lambda\
\eta_E^{(0)}(x) \qquad , \label{eq:ueq21}
\end{equation}
 with parameters
 \[
 \zeta= \frac{2}{\bar{\hbar} } \sqrt{\frac{16}{\bar{\hbar}^2} - 2 E}
 \quad ,
\quad  M^2 =\frac{8}{8 -\bar\hbar^2 E} \quad , \quad
\lambda= M^2 + \frac{4}{\bar{\hbar}^2} (I +
\frac{16}{\bar\hbar^2}) \quad ,
\]
after the change of coordinates: $u={\rm cosh}2 x$.

The $v$-equations (\ref{eq:spec31}), however, become the Mathieu
equations
\begin{equation} - \frac{ d^2 \xi_E^{(0)}(y)}{d y^2} + (
\alpha \, {\rm cos} 4 y + \sigma )\ \xi_E^{(0)}(y) = 0
\label{eq:Mathieu}
\end{equation}
with parameters
\[ \alpha =  \frac{4 E}{\bar{\hbar}^2} \qquad \,\, ,
\qquad \,\, \sigma = \frac{4}{\bar{\hbar}^2}  (I + E
)\qquad ,
\]
under the change: $v={\rm cos}2 y$. The strategy to solve these two entangled equations is
the same as in the case of two centers of different strength.

First, we search for finite solutions of the Razavy equation. The procedure is identical
and we only need to replace $\delta$ by $1$ in the formulae of the previous Section \S. 4.5.
We now have
\[
E_n^{(0)}=\frac{8}{\bar\hbar^2}\left(1-\frac{1}{(n+1)^2}\right) \quad , \quad \zeta_n=\frac{8}{\bar\hbar^2(n+1)}
\quad , \quad I_{nm}=\frac{\bar{\hbar}^2}{4} ( \lambda_{nm} -
(n+1)^2) - \frac{16}{\bar{\hbar}^2}
\]
where $\lambda_{nm}, m=1,2, \cdots ,n+1$ are the roots of the polynomials $P_{n+1}(\lambda)$ that cut the series.
We thus show the results for the lower eigenvalues and eigenfunctions in the next Table:
\begin{table}[htdp]
\begin{center}
\begin{tabular}{|c|c|c|} \hline  Energy & Eigen-function & Separation constant \\ \hline & & \\ $E_0^{(0)}=0$ &
$\eta_{01}^{(0)}(u) = e^{- \frac{4 u}{\bar{\hbar}^2}}$  &
$I_{01}^{(0)} = 0 $\\ \hline & & \\ {\footnotesize
$E_1^{(0)}=\frac{6}{\bar\hbar^2}$} & {\footnotesize $\left\{
\begin{array}{c}
 \eta_{11}^{(0)} (u) = e^{- \frac{2u}{\bar{\hbar}^2}}
\sqrt{2 (u+1)}  \\ \\
\eta_{12}^{(0)}(u) =- e^{- \frac{2 u}{\bar{\hbar}^2}}
\sqrt{2(u-1)}
\end{array}
\right.$} & {\footnotesize $\left\{
\begin{array}{c}
 I_{11}^{(0)} =  -\frac{\bar{\hbar}^2}{4} - \frac{12}{\bar{\hbar}^2} -
 2  \\  \\
I_{12}^{(0)} = -\frac{\bar{\hbar}^2}{4} - \frac{
12}{\bar\hbar^2} + 2
\end{array}
\right.$}
\\ \hline & & \\{\footnotesize $E_2^{(0)}=\frac{64}{
9\bar\hbar^2}$} & {\tiny $\left\{\begin{array}{c}\eta_{21}^{(0)}(u) = - 2 e^{-\frac{4 u}{3\bar{\hbar}^2}} \sqrt{u^2 - 1}  \\ \\
\eta_{22}^{(0)}(u) = \frac{3 \bar{\hbar}^2}{8} e^{-
\frac{4 u}{ 3 \bar{\hbar}^2}} \left[ \frac{16
}{3 \bar{\hbar}^2} u - 1 + \sqrt{ 1 + \frac{256}{ 9
\bar{\hbar}^4}} \right] \nonumber \\ \\
\eta_{23}^{(0)}(u) = \frac{3 \bar{\hbar}^2}{8} e^{-
\frac{4 u}{3 \bar{\hbar}^2}} \left[ \frac{16
}{3 \bar{\hbar}^2} u - 1 - \sqrt{ 1 + \frac{256}{ 9
\bar{\hbar}^4}} \right]\nonumber
\end{array}
\right.$} & {\footnotesize $ \left\{ \begin{array}{c} I_{21}^{(0)} = - \bar{\hbar}^2 - \frac{128}{9 \bar{\hbar}^2} \\ \\
I_{22}^{(0)} = -\frac{\bar{\hbar}^2}{2} - \frac{128}{9
\bar{\hbar}^2} - \frac{1}{6} \sqrt{256 + 9
\bar{\hbar}^4}  \\ \\
I_{23}^{(0)} =  -\frac{\bar{\hbar}^2}{2} - \frac{128}{9
\bar{\hbar}^2} + \frac{1}{6} \sqrt{256 + 9 \bar{\hbar}^4}
\end{array}
\right.$}  \\
\hline
\end{tabular}
\end{center}
\end{table}

Nothing new with respect to the non equal centers case.

\subsubsection{Contribution from the Mathieu equations}

The novelty comes from the Mathieu equations: unlike the Whittaker-Hill equations these equations are never  quasi-exactly solvable -there are no finite solutions whatsoever- but, instead there are lots of solutions that can be described analytically in terms of the Mathieu sine and cosine special functions.

 As in the non-equal centers system, the ground state is an exception. It is not ruled by any Mathieu equation. $\alpha_0=\frac{4}{\bar\hbar^2}E_0=0$,
$\sigma_{01}=\frac{4}{\bar\hbar^2}(I_{01}+E_0)=0$ implies that:
\[
-\frac{d^2\xi_{01}^{(0)}}{d y^2}=0 \, \, \, \, \Rightarrow \, \, \, \, \xi_{01}^{(0)}(y)=A y+B \qquad .
\]
If the two centers have the same strength an important discrete symmetry arises: the $r_1 \leftrightarrow r_2$
exchange is not detectable. $v \leftrightarrow -v$, or, $y \leftrightarrow y+\frac{\pi}{2}$ is a symmetry of the system and we remain with the only invariant solutions $\xi_{01}^{(1)}(y)=B$ under this transformation. There are only two independent choices: $B=1$ and $B=0$. The first choice is \lq\lq even" in $v$, the second choice is \lq\lq odd" in $v$ but negligible.  The zero energy ground state is therefore built from the even $v$-independent wave function: $\psi_{01}^{(0)}(u,v)=\eta_{01}^{(0)}(u)\xi_{01}^{(0)}(v)=  e^{- \frac{4 u}{\bar{\hbar}^2}}$.

The parameters of the Mathieu equations determined by the spectral
problem for positive energy ($n\geq 1$) are:
\[
\alpha_n= 4\frac{E_n} {\bar{\hbar}^2}\qquad \quad , \quad \qquad \sigma_{nm} =4\frac{E_n +
I_{nm}}{\bar{\hbar}^2} \qquad .
\]
The $v \leftrightarrow -v$ symmetry of equation (\ref{eq:spec31}) is translated into the
$y \, \leftrightarrow \, y+\frac{\pi}{2}n$, $n\in{\mathbb Z}$, infinite discrete symmetry of the Mathieu equation.
The Mathieu cosine and Mathieu sine special functions are obtained from the Bloch-type solutions of the Mathieu equation ruled by the discrete translational symmetry with Floquet indices determined from the parameters $\alpha$ and $\sigma$. Because the Mathieu equation is blind to the $v\leftrightarrow -v$ exchange, if $\xi_{nm}^{(0)}(v)$ is a solution $\xi_{nm}^{(0)}(-v)$ also solves (\ref{eq:spec31}). In \cite{MAJM} we chose
even and odd in $v$ combinations of the Mathieu functions, a situation closely related to the hidden quantum supersymmetry in Bosonic systems unveiled in \cite{Plyuschay}. Instead, here we choose the combination
\begin{equation}
 \xi_{nm}^{(0)}(v) =  C\left[ a_{nm},
q_n, \arccos( v ) \right]
+ i  S\left[ a_{nm}, q_n, \arccos( v )
\right] \label{solM} \quad ,
\end{equation}
to build positive energy eigenfunctions, a choice designed to go to the Kepler/Coulomb system at the $d=0$ limit. Here, $C[a,q,z]$ and $S[a,q,z]$ are respectively the cosine and sine Mathieu functions and $q_n =\frac{\alpha_n}{8}$, $a_{nm} = - \frac{\sigma_{nm}}{4}$.

We now list some of the lower values of the parameters
\begin{eqnarray*}
& q_1 = \frac{3}{\bar{\hbar}^4} \qquad , \qquad \left\{
\begin{array}{c} a_{11} = \frac{6}{\bar{\hbar}^4} +\frac{
2}{\bar{\hbar}^2} + \frac{1}{4} \\ a_{12}= \frac{6}{
\bar{\hbar}^4} - \frac{ 2}{\bar{\hbar}^2} + \frac{1}{4}\end{array}\right. \\
& q_2 = \frac{32}{9 \bar{\hbar}^4} \qquad , \qquad \left\{
a_{21} = 1 + \frac{64}{9 \bar{\hbar}^4} \qquad , \qquad
\begin{array}{c} a_{22} =\frac{1}{2} + \frac{64}{9\bar{\hbar}^4} +
\frac{1}{6 \bar{\hbar}^2 } \sqrt{256 + 9 \bar{\hbar}^4} \\
a_{23}= \frac{1}{2} + \frac{64}{9\bar{\hbar}^4} - \frac{1}{
6\bar{\hbar}^2} \sqrt{ 256 + 9 \bar{\hbar}^4}\end{array}\right.
\end{eqnarray*}
and offer a Table showing the probability densities of some eigenfunctions for several values of $\bar\hbar$. For instance, $\bar\hbar=0.7$ is the value of this non-dimensional parameter for the hydrogen
molecule ion. $\bar\hbar=114,7$ corresponds to the ionized helium atom; note that the radius of the nuclei is of
the order of $10^{-15}{\rm cm}$, etcetera. One notices that the smaller $\bar\hbar$ is, the more classical is the system, the wave functions being more concentrated around the centers.

\begin{table}[htdp]
\begin{center}
\caption{ ${\cal N}=2$ supersymmetric wave functions in ${\cal S}{\cal H}_0$ of two equal centers: $\delta=1$} \ \
\begin{tabular}{|c|cccc|}  \hline  & & & & \\
$|\psi_{nm}^{(0)}(x_1,x_2)|^2$ & $\bar{\hbar}=0.7$  & $\bar{\hbar}=1$
& $\bar{\hbar}=2$ & $\bar{\hbar}=4$ \\ & & & &
\\ \hline  & & & & \\ {\footnotesize n=0, m=1 } &
{\includegraphics[height=2cm]{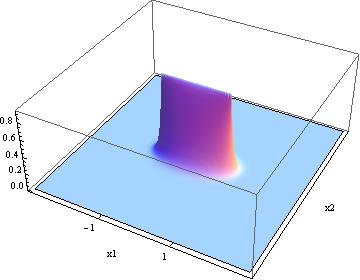}} & {\includegraphics[height=2cm]{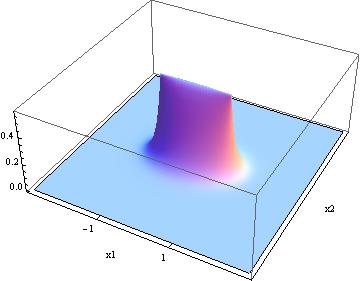}}  &
{\includegraphics[height=2cm]{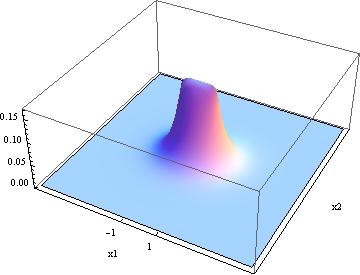}} &
{\includegraphics[height=2cm]{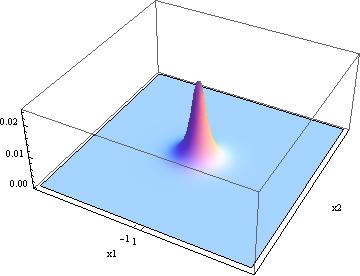}}  \\ & & & & \\
 \hline & & & & \\ {\footnotesize n=1, m=1 } &
{\includegraphics[height=2cm]{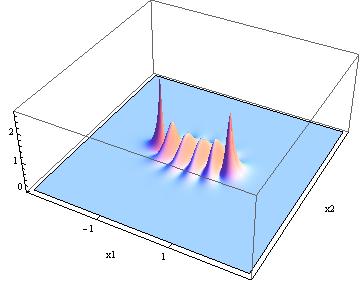}}  &
{\includegraphics[height=2cm]{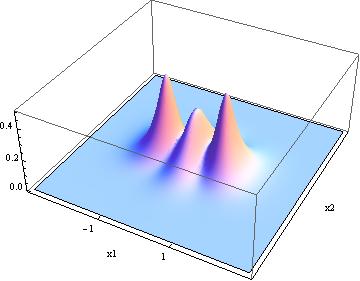}}  &
{\includegraphics[height=2cm]{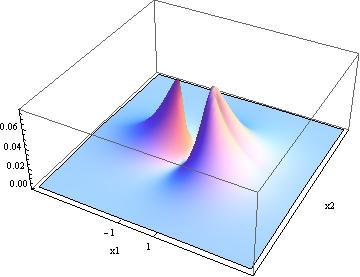}}  &
{\includegraphics[height=2cm]{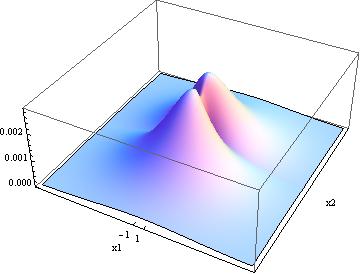}}  \\ & & & & \\
 \hline & & & & \\ {\footnotesize n=1, m=2 } &
{\includegraphics[height=2cm]{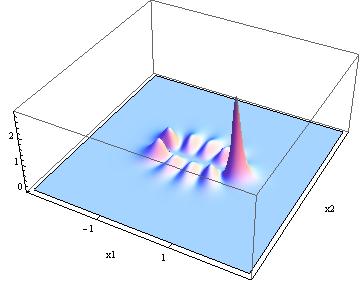}}  &
{\includegraphics[height=2cm]{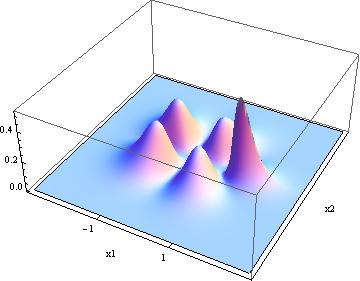}}  &
{\includegraphics[height=2cm]{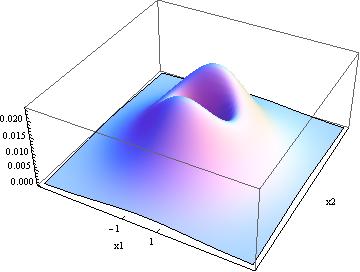}}  &
{\includegraphics[height=2cm]{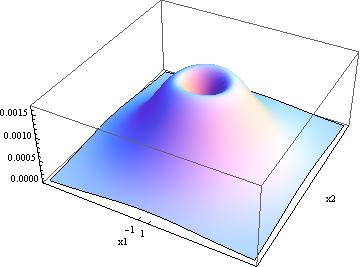}}  \\ & & & & \\
 \hline & & & & \\ {\footnotesize n=2, m=1 } &
{\includegraphics[height=2cm]{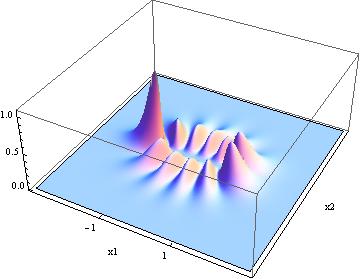}}  &
{\includegraphics[height=2cm]{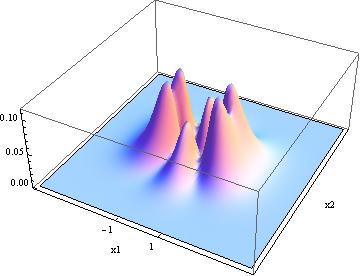}}  &
{\includegraphics[height=2cm]{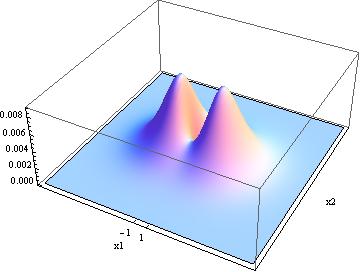}}  &
{\includegraphics[height=2cm]{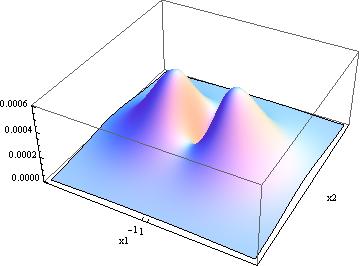}}  \\ & & & & \\
 \hline & & & & \\ {\footnotesize n=2, m=2 } &
{\includegraphics[height=2cm]{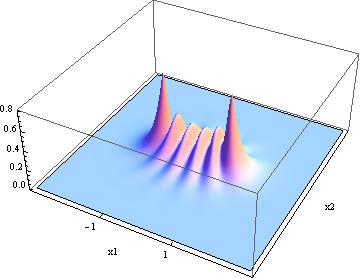}}  &
{\includegraphics[height=2cm]{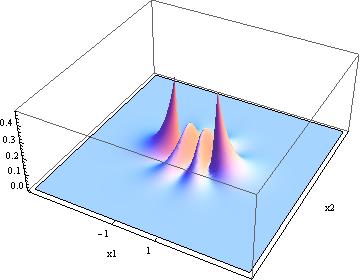}}  &
{\includegraphics[height=2cm]{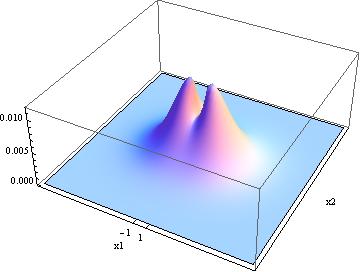}}  &
{\includegraphics[height=2cm]{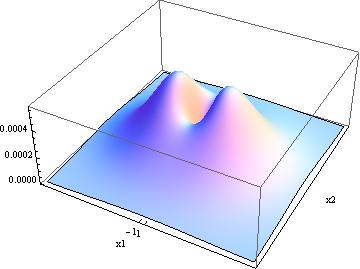}}   \\ & & & & \\
 \hline {\footnotesize n=2, m=3 } &
{\includegraphics[height=2cm]{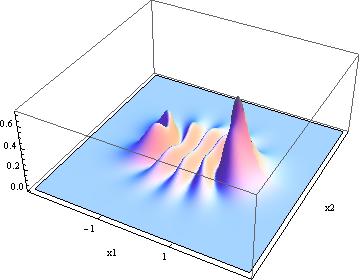}}  &
{\includegraphics[height=2cm]{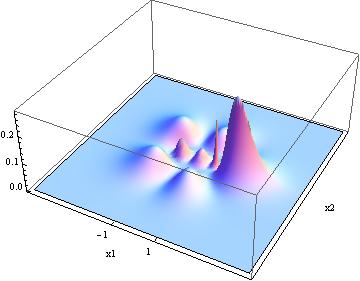}}  &
{\includegraphics[height=2cm]{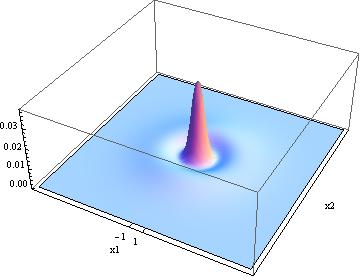}}  &
{\includegraphics[height=2cm]{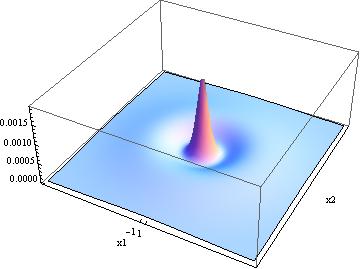}}   \\ & & & & \\
 \hline
 \end{tabular}
\end{center}
\end{table}

\subsection{Comparison between the  ${\cal N}=2$ supersymmetric  and the ${\cal N}=0$
spectra}
The energy spectra of the Euler Hamiltonian - given by the solution of the system of equations
(\ref{scr2c1}), (\ref{scr2c2})- and the $\hat{H}_0$ Euler supersymmetric Hamiltonian
reduced to the ${\cal S}{\cal H}_0$ sector of the supersymmetric Hilbert space -characterized by the system
(\ref{ss2ce1}), (\ref{ss2ce2})-  are listed below:
\[
\begin{array}{ccc}
 \hspace{4.5cm} & \hspace{0.5cm} {\cal N}=0 \qquad \qquad \quad , \qquad \qquad  & {\cal N}=2
\qquad  \\ & \\
\hspace{0.6cm} 0<\delta=\frac{Z_2}{Z_1} \leq 1 \qquad , \qquad &
E_{n}=-\frac{2(1+\delta)^2}{\bar\hbar^2(n+1)^2} \qquad  , \qquad
&
E_{n}^{(0)}=\frac{2(1+\delta)^2}{\bar\hbar^2}\left(1-\frac{1}{(n+1)^2}\right)
\end{array}\quad .
\]
The degeneracy of the $n$th level is $n+1$ in both cases and it is split by the eigenvalues of the symmetry operator,
respectively $\hat{I}$ and $\hat{I}^{(0)}$.

As in the supersymmetric Kepler/Coulomb problem there are bound states in ${\cal S}{\cal H}_1$ of the form:
\[
\psi^{(1)}_{nm}(r_1,r_2)=\hat{Q}^\dagger\, \psi^{(0)}_{nm}(r_1,r_2)\quad , \quad n\geq 1 \, \, , \quad m=1,2, \cdots , n+1 \quad .
\]
Unlike in the supersymmetric Kepler problem, there is one zero energy wave function in ${\cal S}{\cal H}_1$. Coming back to dimension-full coordinates and parameters, it is:
\begin{eqnarray}
\hat{Q}\psi_0^{(1)}(x_1,x_2)&=&\hat{Q}^\dagger\psi_0^{(1)}(x_1,x_2)=0
\qquad , \qquad E^{(1)}_0=0 \nonumber
\\
\psi_{0}^{(1)}(x_1,x_2)&=& \left(
\begin{array}{c} 0\\\mp \frac{r_1+r_2}{4 d r_1 r_2} \sqrt{4 d^2 -(r_1-r_2)^2}
\\ \frac{r_2-r_1}{4 d r_1 r_2} \sqrt{(r_1+r_2)^2-4 d^2 }
 \\ 0
\end{array} \right) e^{- \frac{ 2 m
\alpha (r_2 + \delta r_1)}{\hbar^2}} \label{fzm2} \quad ;
\end{eqnarray}
here, the $\lq \lq-"$ sign occurs for $x_2 > 0$ whereas the $\lq \lq+"$ sign
arises when $x_2<0$. In the limit where the two centers coincide this wave function becomes:
\begin{eqnarray*}\lim_{d\rightarrow 0} \psi_{0}^{(1)}(x_1,x_2)&=&  \left(
\begin{array}{c} 0\\ -\frac{x_2}{x_1^2+x_2^2}
\\ \frac{x_1}{x_1^2+x_2^2}
 \\ 0
\end{array} \right) e^{- \frac{ 2 m
\alpha (1+\delta) }{\hbar^2}\sqrt{x_1^2+x_2^2}}
 \end{eqnarray*}
and we observe that its norm diverges:
\[\lim_{d\rightarrow 0} \int_{-\infty}^\infty\int_{-\infty}^\infty \, dx_1\, dx_2 \, \left|\psi_{0}^{(1)}(x_1,x_2)\right|^2=\lim_{d\rightarrow 0} 2 \pi K_0 \left( \frac{4 m \alpha\, d
}{ \hbar^2} (1+\delta) \right) I_0 \left( \frac{4 m \alpha\, d}{
\hbar^2} (1-\delta) \right) = + \infty \, \, \,  ,
\]
where $K_0(z)$ and $I_0(z)$ are respectively the first-order modified Bessel functions, see \cite{JPA}. Thus,
we confirm that there is no fermionic zero mode in the SUSY Kepler problem. Note that ground states in that
problem must live in the scalar representation of ${\mathbb S}{\mathbb O}(3)$.
\begin{table}[htdp]
\begin{center}
\caption{${\cal N}=2$ supersymmetric  zero modes.} \ \
\begin{tabular}{|c|cccc|}  \hline  & & & & \\
{\footnotesize{Probability density}} & $\bar{\hbar}=0.7$  &
$\bar{\hbar}=1$ & $\bar{\hbar}=2$ & $\bar{\hbar}=4$ \\ & & & &
\\ \hline & & & & \\  ${\delta=\frac{1}{2}\atop|\psi_{01}^{(0)}(x_1,x_2)|^2}$ &
\includegraphics[height=2.2cm]{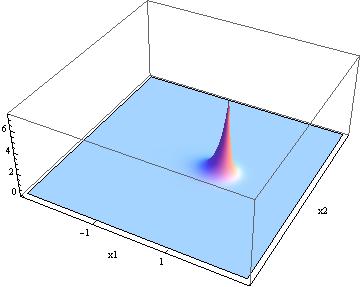} &
\includegraphics[height=2.2cm]{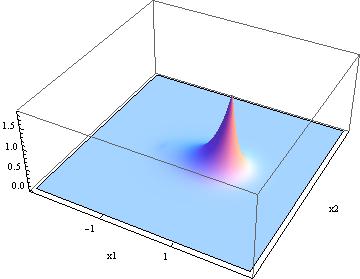} &
\includegraphics[height=2.2cm]{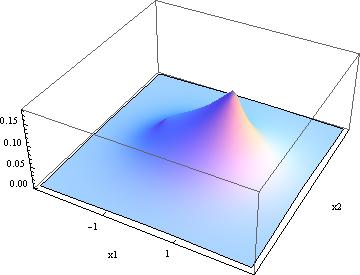} &
\includegraphics[height=2.2cm]{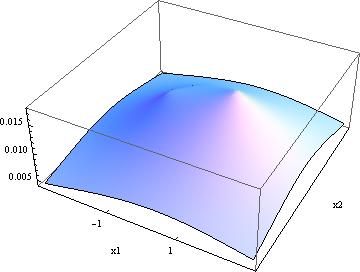}  \\ & & & & \\
 \hline & & & & \\ ${\delta=\frac{1}{2}\atop|\psi_{01}^{(1)}(x_1,x_2)|^2}$&
\includegraphics[height=2.4cm]{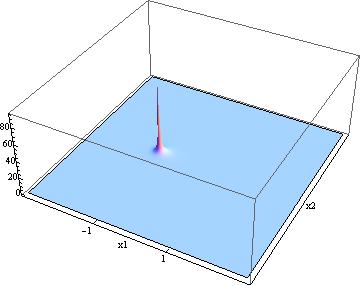}&
\includegraphics[height=2.4cm]{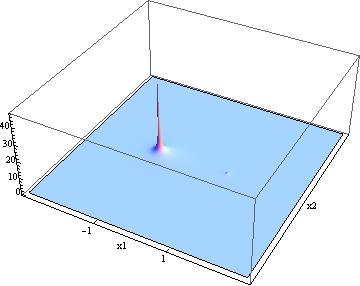} &
\includegraphics[height=2.4cm]{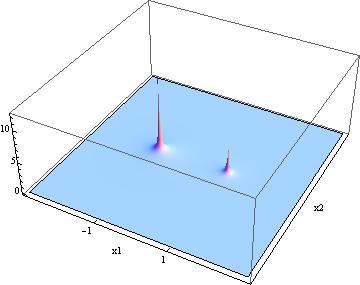} &
\includegraphics[height=2.4cm]{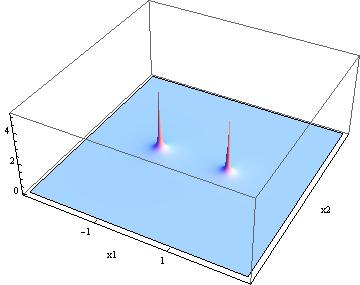}  \\ & & & & \\
\hline & & & & \\ ${\delta = 1\atop |\psi_{01}^{(1)}(x_1,x_2)|^2}$ &
\includegraphics[height=2.4cm]{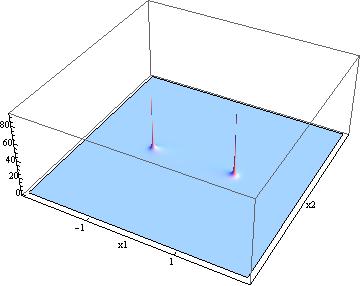}&
\includegraphics[height=2.4cm]{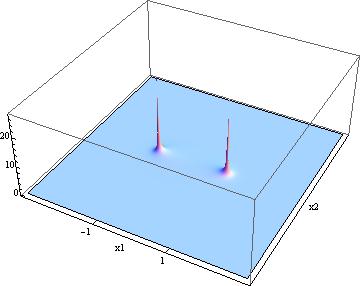} &
\includegraphics[height=2.4cm]{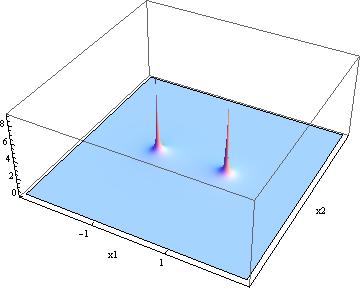} &
\includegraphics[height=2.4cm]{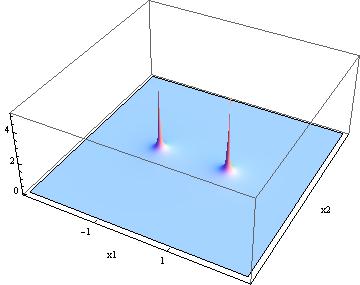} \\ & & & & \\
 \hline
 \end{tabular}
\end{center}
\end{table}

The comparison between the Bosonic and Fermionic zero modes can be
seen in Table 4, where the center on the right is twice as strong as
the center on the left and $\delta=\frac{1}{2}$. In the Bosonic
ground state the electron is concentrated around the stronger center
for small $\bar\hbar$ but becomes spread over the two centers when
$\bar\hbar$ increases. In the Fermionic ground state the
superparticle behaves in the opposite way!: for small $\bar\hbar$ it is
concentrated in the weaker center and two peaks on the two centers
arise for larger $\bar\hbar$. Moreover, for any value of $\bar\hbar$
the probability density of the ground state in ${\cal S}{\cal H}_1$
is always peaked at the centers. In any other respect, i.e.,
concerning the scattering solutions with energy greater than
$\frac{2}{\bar\hbar^2}(1+\delta)^2$, the structure of the spectrum is
qualitatively identical to the spectrum of the supersymmetric Kepler
problem. There are scattering states in ${\cal S}{\cal H}_0$ paired
via the $\hat{Q}^\dagger$ supercharge to scattering states in ${\cal
S}{\cal H}_1$ and scattering states in ${\cal S}{\cal H}_2$ paired
via the $\hat{Q}$ supercharge to scattering states in ${\cal S}{\cal
H}_1$. All this is depicted schematically in Figure 3:

\begin{center}
\begin{figure}[htdp]
\includegraphics[height=6cm]{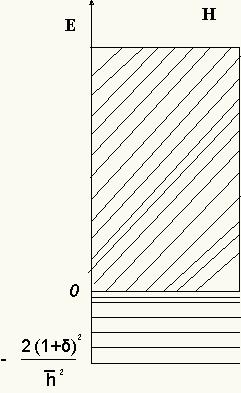}\hspace{1cm} \includegraphics[height=6cm]{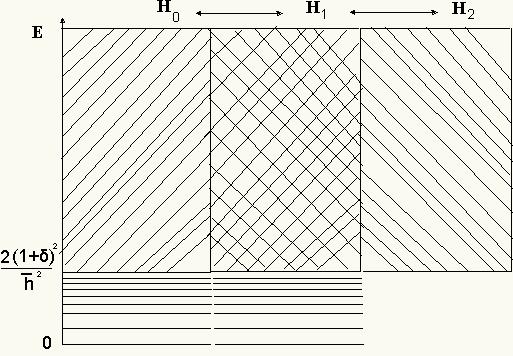}
\caption{The spectrum of the Euler Hamiltonian (left panel). The spectrum of the supersymmetric
Euler Hamiltonian (right panel). }
\end{figure}
\end{center}

\section{Two centers collapse in one center}
In this Section we shall analyze how the two main spectral problems
described respectively in sections \S.3 and \S.4 are connected. The
link appears at the $d=0$ limit of the two-center problem. In order
to go to this singular limit of the modified Euler/Coulomb problem it is
necessary to restore full dimensional variables. Thus, with no
change of notation the parameters and physical variables to be
dealt with in this Section have the proper dimensions. We shall
perform the limit in the case $\delta=1$ because we have full
analytical information for two equal centers.

\begin{itemize}

\item We start from the ground state characterized by: $n=0$ and $m=1$. The eigenvalues of the Hamiltonian and the symmetry operator, the Mathieu parameters, and the ground state wave function, are respectively:
\[
E_0^{(0)} = 0 \quad , \quad I_{01} = 0\quad , \quad q_0 = 0 \quad ,
\quad a_{01} = 0\qquad , \qquad \psi_{01}^{(0)} (u,v) = \left( \frac{m
\alpha}{\hbar^2} \right)\  e^{-\frac{4 m \alpha}{\hbar^2}\, u}
\]
At the $d\rightarrow 0$ limit we have:
\[
r =\lim_{d\to 0} u=\sqrt{x_1^2 + x_2^2} \qquad , \qquad
\varphi=\lim_{d\to 0}{\rm arccos}(\frac{v}{d})={\rm
arctan}(\frac{x_2}{x_1}) \qquad .
\]
Therefore,
\[
\psi_{00}^{(0)} (r,\varphi) = \lim_{d\to 0}\psi_{01}^{(0)} (u, v)=\left(
\frac{m \tilde{\alpha}}{\hbar^2} \right)\  e^{-\frac{2 m
\tilde{\alpha}}{\hbar^2}\, r} \quad , \quad \tilde{\alpha} = 2
\alpha
\]
and the ground state of the one-center problem with twice the
electric charge appears in the $d=0$ limit.

\item We next consider the doublet labeled by $n=1$, $m=1$ and $m=2$. The eigenvalues of the Hamiltonian and symmetry operators are:
\[
E_1^{(0)} = \frac{6 m \alpha^2}{\hbar^2} \qquad , \qquad
\begin{array}{c} I_{11} = -\frac{\hbar^2}{4} - 2 (m d \alpha)
- \frac{12 (m d \alpha)^2}{\hbar^2}  \\ \\ I_{12} =
-\frac{\hbar^2}{4} + 2 (m d \alpha) - \frac{12 (m d
\alpha)^2}{\hbar^2}  \end{array}
\]
The corresponding Mathieu parameters and eigenfunctions read:
\[
q_1 = \frac{ 3 (m d \alpha)^2}{\hbar^4} \qquad , \qquad
\begin{array}{c} a_{11} = \frac{1}{4} + 2 \frac{(m d
\alpha)}{\hbar^2} + \frac{6 (m d \alpha)^2}{\hbar^4}  \\ \\
a_{12} = \frac{1}{4} - 2 \frac{(m d \alpha)}{\hbar^2} + \frac{6 (m d
\alpha)^2}{\hbar^4}  \end{array}
\]
\begin{eqnarray*}
&& \hspace{-0.6cm}\psi_{11}^{(0)} (u,v) = \left( \frac{m
\alpha}{\hbar^2} \right)^{3/2} \,  e^{-\frac{2 m \alpha}{\hbar^2} u}
\, \sqrt{2 (u + d)} \left\{ C[a_{11}, q_1, {\rm arccos}(\frac{v}{d})
+ i S[a_{11}, q_1, {\rm
arccos}(\frac{v}{d})] \right\} \\
&& \hspace{-0.6cm}\psi_{12}^{(0)} (u, v) = - \left( \frac{m
\alpha}{\hbar^2} \right)^{3/2} \,  e^{-\frac{2 m \alpha}{\hbar^2} u}
\, \sqrt{2 (u - d)} \left\{ C[a_{12}, q_1, {\rm arccos}(\frac{v}{d})
+ i S[a_{12}, q_1, {\rm arccos}(\frac{v}{d})] \right\}
\end{eqnarray*}
Because
\begin{eqnarray*}
&& \lim_{d\to 0} q_1 = 0 \qquad , \qquad \alpha_{1k}=\lim_{d\to 0}\,
a_{1k}=\frac{1}{4} \quad , \quad k=1,2 \quad , \quad
\lim_{d\rightarrow 0} I_{11} = \lim_{d\rightarrow 0} I_{12} =
-\frac{\hbar^2}{4}
 \\
&& {\rm cos} \sqrt{\alpha_{1k}}\, \varphi=\lim_{d\to 0} C[a_{1k},
q_1, {\rm arccos}(\frac{v}{d})] \qquad , \qquad {\rm sin}
\sqrt{\alpha_{1k}}\, \varphi=\lim_{d\to 0} S[a_{1k}, q_1, {\rm
arccos}(\frac{v}{d})] \, ,
\end{eqnarray*}
we find
\begin{eqnarray*}
&& \psi_{\frac{1}{2}\frac{1}{2}}^{(0)} (r, \varphi) =\lim_{d\to 0}
\psi_{11}^{(0)} (u, v) = \left( \frac{m \tilde{\alpha}}{\hbar^2}
\right)^{3/2} \ r^{1/2}\ e^{-\frac{ m \tilde{\alpha}}{\hbar^2} r}
\  e^{i \frac{1}{2}\varphi} \\
&& \psi_{\frac{1}{2} -\frac{1}{2}}^{(0)} (r, \varphi) =\lim_{d\to
0}\psi_{12}^{(0)} (u, v)= - \left( \frac{m \tilde{\alpha}}{\hbar^2}
\right)^{3/2} \ r^{1/2} \ e^{-\frac{ m \tilde{\alpha}}{\hbar^2} r} \
e^{-i \frac{1}{2} \varphi} \quad .
\end{eqnarray*}
Again we find that the eigenvalue of the Hamiltonian operator
remains the same at the $d=0$ limit, the eigenvalues of the symmetry
operator go to the eigenvalues of the square of the angular
momentum, and the wave functions become the eigen-functions of the
Kepler/Coulomb problem with spin of one-half. Note that we have
used: $\sqrt{\frac{1}{4}}=\pm\frac{1}{2}$.

\item Finally, we consider the triplet $n=2$, $m=1$, $m=2$ and $m=3$. The eigenvalues of the Hamiltonian and the symmetry operator are:
\[
E_2^{(0)} = \frac{64 m \alpha^2}{ 9 \hbar^2} \qquad , \qquad
\begin{array}{c} I_{21} = -\hbar^2 -\frac{128 (m d \alpha)^2}{9 \hbar^2}  \\ \\ I_{22} =
-\frac{\hbar^2}{2} -\frac{128 (m d \alpha)^2}{ 9 \hbar^2}
-\frac{\hbar^2}{ 2} \sqrt{ 1 + \frac{256 (m d \alpha)^2}{ 9
\hbar^2}}  \\ \\ I_{23} = -\frac{\hbar^2}{2} -\frac{128 (m d
\alpha)^2}{ 9 \hbar^2} + \frac{\hbar^2}{ 2} \sqrt{ 1 + \frac{256 (m
d \alpha)^2}{ 9 \hbar^2}}
\end{array}
\]
whereas the Mathieu parameters read:
\[
q_2 = \frac{ 32 (m d \alpha)^2}{9 \hbar^4} \qquad , \qquad
\begin{array}{c} a_{21} = 1 +  \frac{64 (m d
\alpha)^2}{9 \hbar^4}   \\ \\
a_{22} = \frac{1}{2} +  \frac{64 (m d \alpha)^2}{9 \hbar^4} +
\frac{1}{2} \sqrt{ 1 + \frac{256 (m d \alpha)^2}{9 \hbar^4}} \\ \\
a_{23} = \frac{1}{2} +  \frac{64 (m d \alpha)^2}{9 \hbar^4} -
\frac{1}{2} \sqrt{ 1 + \frac{256 (m d \alpha)^2}{9 \hbar^4}}
\end{array}\quad .
\]
The eigenfunctions are more complicated
\begin{eqnarray*}
&& \psi_{21}^{(0)} (u, v) = \left( \frac{m \alpha}{\hbar^2} \right)^{2}
\,  e^{-\frac{4 m \alpha}{3 \hbar^2} u} \, \sqrt{ u^2 - d^2} \left\{
C[a_{21}, q_2, {\rm arccos}(\frac{v}{d}) + i S[a_{21}, q_2,
{\rm arccos}(\frac{v}{d})] \right\} \\
&& \psi_{22}^{(0)} (u, v) = \left( \frac{m \alpha}{\hbar^2} \right)^{2}
\frac{ 3 \hbar^2}{8 m \alpha} \, e^{-\frac{4 m \alpha}{3 \hbar^2} u}
\, \left[ \frac{ 16 m \alpha}{3 \hbar^2} u - 1 + \sqrt{1 + \frac{256
(m d \alpha)^2}{9 \hbar^2} } \right] \\ && \qquad \qquad \qquad
\times \left\{ C[a_{22}, q_2, {\rm arccos}(\frac{v}{d}) + i
S[a_{22}, q_2, {\rm arccos}(\frac{v}{d})] \right\}
\\
&& \psi_{23}^{(0)} (u, v) = \left( \frac{m \alpha}{\hbar^2} \right)^{2}
\frac{ 3 \hbar^2}{8 m \alpha} \, e^{-\frac{4 m \alpha}{3 \hbar^2} u}
\, \left[ \frac{ 16 m \alpha}{3 \hbar^2} u - 1 - \sqrt{1 + \frac{256
(m d \alpha)^2}{9 \hbar^2} } \right] \\ && \qquad \qquad \qquad
\times \left\{ C[a_{23}, q_2, {\rm arccos}(\frac{v}{d}) + i
S[a_{23}, q_2, {\rm arccos}(\frac{v}{d})] \right\}
\end{eqnarray*}
but the limits are similar:
\begin{eqnarray*}
&& \lim_{d\to 0} q_2 = 0 \qquad , \qquad \alpha_{21}=\lim_{d\to 0}\,
a_{21}=1=\lim_{d\to 0}\, a_{22}=\alpha_{22} \quad , \quad
\alpha_{23}=\lim_{d\to 0}\, a_{23}=0 \\ && \lim_{d\rightarrow 0}
I_{21} = \lim_{d\rightarrow 0} I_{22} = -\hbar^2 \quad , \quad
\lim_{d\rightarrow 0} I_{23} = 0 \quad , \, \, k=1,2,3
 \\
&&\hspace{-0.8cm} {\rm cos} \sqrt{\alpha_{2k}}\, \varphi=\lim_{d\to
0} C[a_{2k}, q_2, {\rm arccos}(\frac{v}{d})] \qquad , \qquad {\rm
sin} \sqrt{\alpha_{2k}}\, \varphi=\lim_{d\to 0} S[a_{2k}, q_2, {\rm
arccos}(\frac{v}{d})]\, \, .
\end{eqnarray*}
Thus, we find
\begin{eqnarray*}
&& \psi_{11}^{(0)} (r,\varphi) = \lim_{d\to 0}\psi_{21}^{(0)} (u,v)=\left(
\frac{m \tilde{\alpha}}{\hbar^2} \right)^{2} \ r\ e^{-\frac{2 m
\tilde{\alpha}}{3 \hbar^2} \bar{r}}
\  e^{i  \varphi} \\
&& \psi_{1, -1}^{(0)} (r, \varphi) = \lim_{d\to 0}\psi_{22}^{(0)} (u,v)=
\left( \frac{m \tilde{\alpha}}{\hbar^2} \right)^{2} \ r \
e^{-\frac{2 m \tilde{\alpha}}{3 \hbar^2} r} \ e^{-i  \varphi}
\\
&& \psi_{1 0}^{(0)} (r, \varphi) = \lim_{d\to 3}\psi_{23}^{(0)} (u,v)=
\left( \frac{m \tilde{\alpha}}{\hbar^2} \right)^{2} \ \left( \frac{3
\hbar^2}{ m\tilde{\alpha}} - 4 r \right) \ e^{-\frac{2 m
\tilde{\alpha}}{3 \hbar^2} r}
\end{eqnarray*}
again falling in wave functions of one doubly charged center,
in this case with angular momentum 1.
\end{itemize}
3D graphics of this analysis are shown in Table 5. We remark that the
graphics are drawn in non-dimensional variables and some of them
cannot skip the 1 in the $d=0$ limit displaying non-smooth tendency
to the Kepler wave functions; namely the $n=2,m=1$ and $n=2,m=2$
wave functions.

\begin{table}[htdp]
\begin{center}
\caption{3D Plots of the probability densities: Kepler versus
Euler.}
\begin{tabular}{|c|cccc|}  \hline  & & & & \\
{\footnotesize{Probability}} &  &  &  & \\
{\footnotesize{density}} & & & & \\
\hline & & & &  \\ $ \hbar=1$  & $j = 0$, $m=0$ & $j=\frac{1}{2}$,
$m=\pm\frac{1}{2}$ & $j=1$, $m=\pm 1$ & $j=1$, $m=0$ \\
\hline & & & &  \\
{\footnotesize{Kepler}}    &
\includegraphics[height=2.2cm]{figura2a.jpg} &
\includegraphics[height=2.2cm]{figura2b.jpg} &
\includegraphics[height=2.2cm]{figura4a.jpg} &
\includegraphics[height=2.2cm]{figura4b.jpg}  \\ & & & & \\ & & & &
\\ \hline & & & &  \\ $ \bar{\hbar} >> 1$ & $n = 0$, $m=1$ & $n=1$, $m=1$ & $n=2$, $m=1$ & $n=2$, $m=3$
\\ \hline & & & &  \\{\footnotesize{Two Centers}}  &
\includegraphics[height=2.1cm]{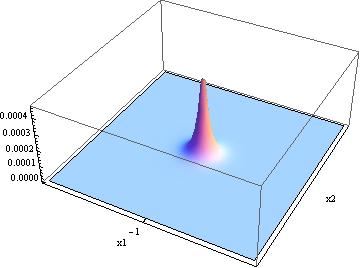}&
\includegraphics[height=2.1cm]{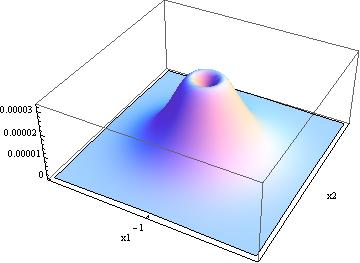} &
\includegraphics[height=2.1cm]{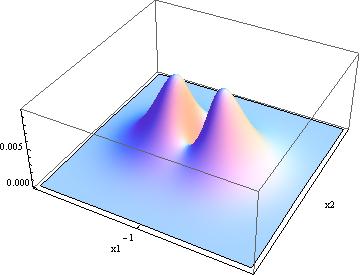} &
\includegraphics[height=2.1cm]{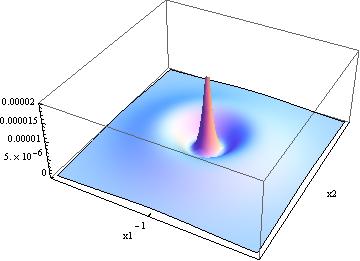}  \\ & & & & \\ & & & &
\\ \hline & & & &  \\ $ \bar{\hbar} >> 1$  &  & $n=1$, $m=2$ & $n=2$, $m=2$ &
\\ \hline & & & &  \\ {\footnotesize{Two Centers}}   & $ \ $ &
\includegraphics[height=2.1cm]{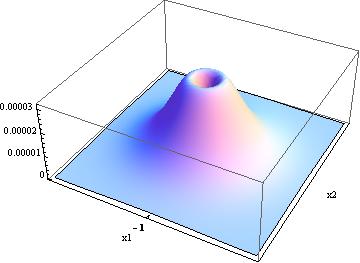} &
\includegraphics[height=2.1cm]{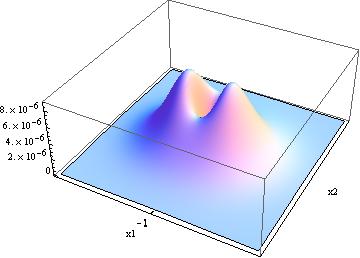} & $ \ $
 \\ & & & & \\
 \hline
 \end{tabular}
\end{center}
\end{table}

\subsection{Collapse of two centers of different strength}
When the two centers have different strengths, $\delta\neq 1$, the one center collapse happens exactly in the same way
as compared to the equal two-center collapse because $\lim_{d=0}B_{nm}=\lim_{\delta=1}B_{nm}=0$. Id est, when the charges are superposed the only things that matter is the total charge. We now offer this analysis for completeness
which will shed light on the physical nature of the recurrence relations (\ref{recr}) and the basic solutions (\ref{bvsol}) of the Whittaker-Hill equations. Because we prefer to deal with power series with purely numerical coefficients, we shall stick to non-dimensional $v$-variables and put all the physical dimensions back in the $u$-dependent part of the wave functions. We recall that:
\[
A_{nm}=-\frac{1}{\bar\hbar^2}\left(I_{nm}+\frac{4(1-\delta)^2}{\bar\hbar^2}\right)\quad , \quad B_{nm}=\frac{2(1-\delta)}{\bar\hbar^2} \quad , \quad C_{nm}=-\frac{2}{\bar\hbar^2}\left(E_n-\frac{2(1-\delta)}{\bar\hbar^2}\right)
\]
are the non-dimensional parameters of the algebraic WH equations.
\begin{itemize}

\item Start with the ground state: $n=0$, $m=1$. The values $A_{01}=-\left(\frac{2(1-\delta)}{\bar\hbar^2}\right)^2$, $B_{01}=\frac{2(1-\delta)}{\bar\hbar^2}$, and
    $C_{01}=\frac{4(1-\delta)^2}{\bar\hbar^4}$ means that the recurrence relations (\ref{recr}) are solved
    by $c_{01}^{(k)}=\frac{1}{k!}\left(\frac{2(1-\delta)}{\bar\hbar^2}\right)^k$. Thus, $\xi_{01}^{(0)}(v)=\sum_{k=0}^\infty\, \frac{1}{k!}\left(\frac{2(1-\delta)}{\bar\hbar^2}\right)^k v^k$ must be multiplied by the unique finite solution of the $E_0=I_{01}=0$ Razavy equation:
\[
\psi_{01}^{(0)} (u, v) = \left( \frac{m \alpha
(1+\delta)}{\hbar^2} \right)\  e^{-\frac{2 m \alpha (1 + \delta)
}{\hbar^2}\ u} \  e^{\frac{2 m \alpha d (1 -\delta) }{\hbar^2}\
v} \quad .
\]
It is clear, like above, that now
\[
\lim_{d=0}\psi_{01}^{(0)} (u, v) =
\left( \frac{m
\tilde{\alpha}}{\hbar^2} \right)\  e^{-\frac{2 m
\tilde{\alpha}}{\hbar^2}\ r}=\psi_{00}^{(0)} (r, \varphi)  \quad , \quad \tilde{\alpha}=(1+\delta)\alpha
\]
and we find the ground state of only one-center with zero energy and momentum and electric charge $(1+\delta)\alpha$.

\item Next, we consider the doublet: $n=1$ \, , $\left\{\begin{array}{c} m=1 \\ m=2 \end{array}\right.$.
The eigenfunctions are of the form
\begin{eqnarray*}
&& \psi_{11}^{(0)} (u, v) = \left( \frac{m \alpha (1 +
\delta)}{\hbar^2} \right)^{3/2} \,  e^{-\frac{ m (1+\delta)
\alpha}{\hbar^2} u} \, \sqrt{2 (u + d)}\  \xi_{11}^{(0)} (v) \\
&& \psi_{12}^{(0)} (u, v) = - \left( \frac{m \alpha (1
+ \delta)}{\hbar^2} \right)^{3/2} \,  e^{-\frac{ m (1 + \delta)
\alpha}{\hbar^2} u} \, \sqrt{2 (u - d)} \
\xi_{12}^{(0)}(v)
\end{eqnarray*}
where the $\xi_{11}^{(0)}(v)$ and $\xi_{12}^{(0)}(v)$ are linear combinations of the basic solutions of the WH equations to be identified in such a way that the Kepler/Coulomb wave functions of energy $E_1=\frac{3m\tilde{\alpha}^2}{\hbar^2}$ and angular momentum $\frac{\hbar}{2}$ and $-\frac{\hbar}{2}$ arise at $d=0$.

The pass to the limit is identical to the collapse of two equal centers in the $u$-part of the wave function. Therefore, we will describe in detail only the collapse of the $v$-part.
Because the $d=0$ limits of the WH parameters are
\[
\lim_{d\to 0}A_{11}=\lim_{d\to 0}A_{12}=\frac{1}{4} \quad , \quad \lim_{d\to 0}B_{11}=\lim_{d\to 0}B_{12}=0=\lim_{d\to 0}C_{11}=\lim_{d\to 0}C_{12}
\]
the recurrence relations become very easy: if we denote $q=1,2$,
\[
c_{1q}^{(2)}=-\frac{1}{8}c_{1q}^{(0)} \quad , \quad
c_{1q}^{(3)}=-\frac{1}{6}\left(\frac{1}{4}-1\right)c_{1q}^{(0)} \quad ,
\quad
c_{1q}^{(k)}=-\frac{1}{k(k-1)}\left(\frac{1}{4}-(k-2)^2\right)c_{1q}^{(k-2)}
\quad .
\]
The basic solutions, corresponding respectively to the choices $c^{(0)}_{1q+}=1, c^{(1)}_{1q+}=0$ and $c^{(0)}_{1q-}=0, c^{(1)}_{1q-}=1$, are:
\begin{eqnarray*}
&& c^{(2)}_{1q+}=-\frac{1}{8} \, \, \, , \, \, \, c^{(4)}_{1q+}=-\frac{5}{128} \, \, \, , \, \, \, c^{(6)}_{1q+}=\frac{21}{1024}\, , \cdots \quad , \quad \xi^{(0)}_{1q+}(v)=\sum_{k=0}^\infty \, c^{(2k)}_{1q+}v^{2k} \\
&& c^{(3)}_{1q-}=\frac{1}{8} \, \, \, , \, \, \, c^{(5)}_{1q-}=\frac{7}{128} \, \, \, , \, \, \, c^{(7)}_{1q-}=\frac{33}{1024}\, , \cdots \quad , \quad \xi^{(0)}_{1q-}(v)=\sum_{k=0}^\infty \, c^{(2k+1)}_{1q+}v^{2k+1}\quad .
\end{eqnarray*}
We know that $C[1/4,0,{\rm arccos}v]\pm i S[1/4,0,{\rm arccos}v]$ goes to ${\rm exp}[\pm i\frac{\varphi}{2}]$ at $d=0$. Comparison with the power series expansion
\begin{eqnarray*}
&& C[1/4,0,{\rm arccos}v]+ i S[1/4,0,{\rm arccos}v]=\\&=&\frac{1+i}{\sqrt{2}}+\frac{\left(\frac{1}{2}-\frac{i}{2}\right) v}{\sqrt{2}}-\frac{\left(\frac{1}{8}+\frac{i}{8}\right)
   v^2}{\sqrt{2}}+\frac{\left(\frac{1}{16}-\frac{i}{16}\right) v^3}{\sqrt{2}}-\frac{\left(\frac{5}{128}+\frac{5 i}{128}\right)
   v^4}{\sqrt{2}}\\ &+&\frac{\left(\frac{7}{256}-\frac{7 i}{256}\right) v^5}{\sqrt{2}}-\frac{\left(\frac{21}{1024}+\frac{21 i}{1024}\right)
   v^6}{\sqrt{2}}+O\left(v^7\right)\quad ,
\end{eqnarray*}
and the analogous series with the relative minus sign, tells us that the right combinations such that $\lim_{d\to 0}\xi_{11}^{(0)}(v)={\rm exp}[i\frac{\varphi}{2}]$ and $\lim_{d\to 0}\xi_{12}^{(0)}(v)={\rm exp}[-i\frac{\varphi}{2}]$ are:
\[
\xi^{(0)}_{1q}(v)=\frac{1-i(-1)^q}{\sqrt{2}}\xi^{(0)}_{1q+}+\frac{1+i(-1)^q}{2\sqrt{2}}\xi^{(0)}_{1q-} \quad .
\]
Henceforth,
\begin{eqnarray*}
&& \lim_{d\to 0}\psi_{11}^{(0)} (u, v)=\lim_{d\to 0}\eta^{(0)}_{11}(u)\xi_{11}^{(0)}(v)=
 \left( \frac{m \tilde{\alpha}}{\hbar^2} \right)^{3/2} \
r^{1/2}\ e^{-\frac{ m \tilde{\alpha}}{\hbar^2} r}
\  e^{i \frac{1}{2} \varphi}=\psi_{\frac{1}{2}\frac{1}{2}}^{(0)} (r, \varphi) \\
&& \lim_{d\to 0}\psi_{12}^{(0)} (u, v)=\lim_{d\to 0}\eta_{12}^{(0)}(u,v)\xi_{12}^{(0)}(v)= -
\left( \frac{m \tilde{\alpha}}{\hbar^2} \right)^{3/2} \
r^{1/2} \ e^{-\frac{ m \tilde{\alpha}}{\hbar^2} r} \
e^{-i \frac{1}{2} \varphi}=\psi_{\frac{1}{2} -\frac{1}{2}}^{(0)} (r, \varphi) \quad .
\end{eqnarray*}

\item Finally,  we address the triplet state $n=1$ \, , $\left\{\begin{array}{c}m=1 \\ m=2 \\ m=3\end{array}\right.$.
The $d=0$ limits of the WH parameters are in this case
\[
\lim_{d\to 0}A_{21}=\lim_{d\to 0}A_{22}=1 \, \, , \, \, \lim_{d\to 0}A_{23}=0 \quad , \quad \lim_{d\to 0}B_{2q}=0=\lim_{d\to 0}C_{21q} \, \, , \quad q=1,2,3 \quad .
\]
The basic solutions for $q=1,2$ are:
\begin{eqnarray*}
&& c_{2q+}^{(0)}=1 \, \, , \, \, c_{2q+}^{(1)}=0 \, \, , \, \, c^{(2)}_{2q+}=-\frac{1}{2} \, \,  ,
\, \, c^{(4)}_{2q+}=-\frac{1}{8}  \, \, , \, \,  c^{(6)}_{2q+}=-\frac{1}{16}\, \, , \, \, c^{(8)}_{2q+}=
-\frac{5}{128}\cdots \\
&& c^{(0)}_{2q-}=0 \, \, , \, \, c^{(1)}_{2q-}=1 \, \, , \, \,
c^{(k)}_{2q-}=0 \, \, \, , \, \forall k\geq 2\, , \cdots \\ &&
\xi^{(0)}_{2q+}(v)=\sum_{k=0}^\infty \, c^{(2k)}_{1q+}v^{2k}  \qquad
, \qquad \xi^{(0)}_{2q-}(v)=\sum_{k=0}^\infty \,
c^{(2k+1)}_{1q+}v^{2k+1}=v \quad  \qquad .
\end{eqnarray*}
Comparison with the power series expansion
\[
C[1,0,{\rm arccos}v]+i S[1,0,{\rm arccos}v]=i+v-\frac{i v^2}{2}-\frac{i v^4}{8}-\frac{i v^6}{16}-\frac{5 i v^8}{128}-\frac{7 i v^{10}}{256}+O\left(v^{11}\right)
\]
show us that the combinations
\[
\xi^{(0)}_{21}(v)=i\xi^{(0)}_{21+}(v)+\xi^{(0)}_{21-}(v) \qquad , \qquad \xi^{(0)}_{22}(v)=-i\xi^{(0)}_{22+}(v)+\xi^{(0)}_{22-}(v)
\]
go respectively to ${\rm exp}[i\varphi]$ and ${\rm exp}[-i\varphi]$ at the $d=0$ limit.

The recurrence relations for $n=2$, $m=3$ are solved by
\begin{eqnarray*}
&&\hspace{-0.7cm} c^{(0)}_{23+}=1 \, \, , \, \, c^{(1)}_{23+}=0 \, \, , \, \, c^{(k)}_{23+}=0 \, \, , \, \forall k\geq 2 \quad , \quad \xi^{(0)}_{23+}(v)=1 \\ &&\hspace{-0.7cm} c^{(0)}_{23-}=0 \, \, , \, \, c^{(1)}_{23-}=1 \, \, , \, \, c^{(3)}_{23-}=\frac{1}{6} \, \, , \, \, c^{(5)}_{23-}=\frac{3}{40} \, \, , \, \, \, c^{(7)}_{23-}=\frac{5}{112} , \cdots \, \xi^{(0)}_{23-}(v)=\sum_{k=0}^\infty \, c^{(2k+1)}_{1q+}v^{2k+1} \, \, .
\end{eqnarray*}
Because $C[0,0,{\rm arccos}v]+i S[0,0,{\rm arccos}v]=\frac{1}{\sqrt{2}}$ the linear combination $\xi^{(0)}_{23}(v)=\frac{1}{\sqrt{2}}\xi^{(0)}_{23+}(v)$ leads to the Kepler/Coulomb eigenfunction in the $d=0$ limit.

In fact, we obtain in this limit the right Kepler/Coulomb triplet with energy $E_2 ^{(0)}= \frac{16 m \tilde{\alpha}}{9 \hbar^2} $ and angular momenta 1 -$
\lim_{d\rightarrow 0} I_{21} = \lim_{d\rightarrow 0}
I_{22} = -\hbar^2 $, $\lim_{d\rightarrow 0} I_{23} = 0$-:
\begin{eqnarray*}
&& \lim_{d\to 0}\psi_{21}^{(0)} (u,v)=\lim_{d\to 0}\eta^{(0)}_{21}(u)\xi^{(0)}_{21}(v)
 = \left(\frac{m
\tilde{\alpha}}{\hbar^2} \right)^{2} \ r\ e^{-\frac{2 m
\tilde{\alpha}}{3 \hbar^2} r}
\  e^{i  \varphi}=\psi_{11}^{(0)} (r,\varphi) \\
&& \lim_{d\to 0}\psi_{22}^{(0)} (u,v)=\lim_{d\to 0}\eta^{(0)}_{22}(u)\xi^{(0)}_{22}(v)  = \left( \frac{m
\tilde{\alpha}}{\hbar^2} \right)^{2} \ r \ e^{-\frac{2 m
\tilde{\alpha}}{3 \hbar^2} r} \ e^{-i\varphi}=\psi_{1,
-1}^{(0)} (r, \varphi)
\\
&& \lim_{d\to 0}\psi_{23}^{(0)} (u,v)=\lim_{d\to 0}\eta^{(0)}_{23}(u)\xi^{(0)}_{23}(v)  = \left( \frac{m
\tilde{\alpha}}{\hbar^2} \right)^{2} \ \left( \frac{3 \hbar^2}{
m\tilde{\alpha}} - 4 r \right) \ e^{-\frac{2 m
\tilde{\alpha}}{3 \hbar^2} r}=\psi_{1
0}^{(0)} (r, \varphi)\qquad .
\end{eqnarray*}

\end{itemize}

\section{Further comments}
We end this long survey by thinking about further possible extensions of these ideas and calculations to other
classical integrable systems.
\begin{enumerate}

\item The immediate temptation is to address the Kepler/Coulomb problem in ${\mathbb R}^N$. The $N$-dimensional supersymmetric Kepler/Coulomb problem has already been developed and fully solved in \cite{KLPW}. In our approach it would be easily doable in the $\hat{N}=0$/$\hat{N}=1$ and $\hat{N}=N$/$\hat{N}=N-1$ sectors; the wave functions in $L^2({\mathbb R}^+\times {\mathbb S}^{N-1})$ would be organized in irreducible representations of ${\mathbb S}{\mathbb O}(N+1)$ -the bound states- or ${\mathbb S}{\mathbb O}(N,1)$ -the scattering states-. In both cases, the wave functions are Kummer confluent functions times the spherical harmonics in $L^2({\mathbb S}^{N-1})$. Of course, the structure of the supersymmetric Hilbert space is more complicated; there are $N+1$ sub-spaces of dimension $\left(\begin{array}{c} N \\ k\end{array}\right), k=0,1, \cdots N$, so that: $\sum_{k=0}^N \, \left(\begin{array}{c} N \\ k\end{array}\right)=2^N$. This means
    that in the $k$-subspaces such that $2\leq k \leq N-2$ only the procedure proposed in \cite{KLPW} would be effective.

    The supersymmetric modified Euler/Coulomb problem in ${\mathbb R}^N$ does not pose more difficulties than those encountered in this paper because all the $N-2$ additional variables are cyclic.

\item It will also be interesting to build the supersymmetric extension of the Kepler/Coulomb problem constrained
to a sphere. This problem in the non supersymmetric framework was addressed independently by Higgs and Pronko in \cite{Higgs} and \cite{Pronko}. The supersymmetric generalization will require to deal with all the subtleties of spinors living in non-flat manifolds. The metric, the $N$-bein, the spin connection, and the like will enter the supercharges one way or another to make the system more intricate, see e.g. \cite{Ioffe8} where supersymmetric systems in curvilinear coordinates are constructed.

The Euler problem considered on a sphere, see \cite{Vosmis}, is also a fairly well known integrable system with
applications in celestial mechanics. It seems promising and interesting to work out the corresponding supersymmetric extension.

\item Another important integrable system is the Neumann problem \cite{Neumann}: a particle forced to move on a sphere under the action of attractive elastic forces. This has been a source of inspiration for treatises on dynamical integrable systems, see \cite{Moser} and \cite{Dubrovin}, and it has been applied, in the repulsive case, by some of us to study topological defects in non-linear sigma models with quadratic \cite{AMAJ} and  quartic \cite{AMAJ1} potentials.
    We believe that the supersymmetric extension of the quantum version of the Neumann problem will provide a physical example of the systems envisaged by Witten in \cite{Wit3} to construct a quantum derivation of Morse theory.

\item The Bosonic zero modes $\psi^{(0)}_{00}(r,\phi)$ in (\ref{hwwf}) and $\psi^{(0)}_{01}(u,v)$ in (\ref{bzm2}) have been easy to find. The Fermionic zero mode $\psi^{(1)}_0(x_1,x_2)$ (\ref{fzm2}) in the two center problem was discovered by means of supercharges defined in elliptic coordinates, translated back to Cartesian coordinates, and checked within a Mathematica environtment. A Fermionic zero mode in the one center problem, however, does not exist \cite{Wipf}, \cite{KLPW}. This is very intriguing and compels us to study SUSY quantum mechanics in curvilinear coordinates in a more profound way. The way forward to and backward from the curvilinear to Cartesian coordinates of all these structures is highly non-trivial, see \cite{Ioffe8} for early attempts in this program. We think that the difficulties with the zero modes have a similar origin to the subtleties arising in the definition of spinors on curved manifolds. We plan to analyze this issue in a future publication.

\end{enumerate}

\section*{Acknowledgements}

 All of us warmly acknowledge Mikhail Ioffe for lectures, seminars, talks, and conversations
 on Supersymmetric Quantum Mechanics over recent years. His mastery of supersymmetric QM has greatly helped us in the struggle to improve our understanding of a matter with so many facets.

 JMG is also indebted to Andreas Wipf for patiently hearing from him about the two centers problem and for the
 crucial suggestion of studying the Euler problem in close comparison with the Kepler problem. Besides being a very fruitful idea, it showed us how to shape the structure of the paper.

We also thank Primitivo Acosta-Humanez, David Blazquez, Mayerling Nu$\tilde{\rm n}$ez-Portela and Mikhail Plyuschay for giving us the opportunity to present this material at the Jairo Charris Seminar 2010 {\it Algebraic aspects of Darboux transformations, quantum integrable systems, and supersymmetric quantum mechanics}, Santa Marta, Colombia, 4-7 August 2010. Part of this material was presented almost immediately before in the Workshop {\it Supersymmetric quantum mechanics and quantum spectral design}, 18-28 July 2010, Benasque, Spain. We warmly thank the Benasque organizers Alexander Andrianov, Luismi Nieto, and Javier Negro as well for their kind invitation to participate. The two workshops, from the Aneto peak in the Pyrenees to the Sierra Nevada de Santa Marta on the Caribbean coast, were indeed very stimulating meetings on closely related subjects celebrated in rapid succession !!

Finally, we gratefully acknowledge that this work has been partially financed
by the Spanish Ministerio de Educacion y Ciencia (DGICYT) under grant: FIS2009-10546.

\end{document}